\documentclass[12pt,preprint]{aastex}           
\usepackage[usenames,dvipsnames]{color}

\begin{document}

\title{Wind structure and luminosity variations in  the WR/LBV HD 5980\footnote{Based on data obtaned with 
HST, IUE, FUSE and the 6.5-m Magellan telescopes at Las Campanas Observatory in Chile}}

\author{Leonid Georgiev\altaffilmark{1}}
\affil{Instituto de Astronom\'{\i}a, Universidad Nacional Aut\'onoma de M\'exico,
Apdo. Postal 70-264, M\'exico D.F., 04510, georgiev@astro.unam.mx}

\author{Gloria Koenigsberger}
\affil{Instituto de Ciencias F\'{\i}sicas, Universidad Nacional Aut\'onoma de M\'exico,
Apdo. Postal 48-3, Cuernavaca, Mor. 62210, gloria@astro.unam.mx}

\author{D. John Hillier}
\affil{Department of Astronomy, 3941 O'Hara Street, University of Pittsburg, Pittsburg, PA 15260, USA}

\author{Nidia Morrell}
\affil{Las Campanas Observatory, The Carnegie Observatories,
Colina El Pino s/n, Casillas 601, La Serena, Chile}

\author{Rodolfo Barb\'a}
\affil{Departamento de F\'{\i}sica, Universidad de la Serena, Benavente 980,
La Serena, Chile; ICATE-CONICET, San Juan Argentina}
\and
\author{Roberto Gamen}
\affil{Facultad de Ciencias Astron\'omicas y Geof\'{\i}sicas, Universidad Nacional de La Plata,
and Instituto de Astrof\'{\i}sica de La Plata (CCT La Plata-CONICET), Paseo del
Bosque S/N, B1900FWA, La Plata, Argentina}

\altaffiltext{1}{Visiting Astronomer, Department of Astronomy, 3941 O'Hara Street, University of Pittsburg, Pittsburg, PA 15260, USA}

\footnote{Based on data obtaned with HST, IUE, FUSE and the 6.5-m Magellan telescopes at Las
Campanas Observatory in Chile}

\begin{abstract}

Over the past 40 years, the massive LBV/WR  system HD 5980 in the Small Magellanic Cloud has 
undergone  a long-term S Doradus type variability cycle and two brief and violent eruptions in 1993 and 1994.
In this paper we  analyze a collection of UV  and optical spectra obtained between 1979 and 2009 and
perform CMFGEN model fits to spectra of 1994, 2000, 2002 and 2009.  The results are as follows: 
a) The long term S Dor-type variability is associated with changes of the hydrostatic radius; 
b) The  1994 eruption involved changes in its bolometric luminosity and wind structure;
c) the emission-line strength, the wind velocity and the continuum luminosity underwent  correlated 
variations in the sense that a decreasing V$_\infty$ is associated with  increasing  emission line and
continuum levels;  and d) The spectrum of the third star in the system ({\it Star C}) is well-fit by a 
T$_{eff}$=32 K model atmosphere with SMC chemical abundances. 

For all epochs, the  wind of the erupting star is optically thick at the sonic point and is thus driven mainly 
by the continuum opacity.  We speculate that the wind switches between two stable regimes 
driven by the ``hot" (during the eruption) and the ``cool" (post-eruption) iron opacity bumps as defined 
by Lamers \& Nugis (2002) and Gr\"afener and Hamann (2008), and thus the wind may undergo a bi-stability jump 
of a different nature  from that which occurs in OB-stars.  

\end{abstract}


\section{Introduction}

Eruptive mass-loss phenomena  in massive stars is emerging as an area of
interest for many reasons, one of the most important of which is that the
characteristics of certain supernova explosions depend critically on the progenitor's mass.
The stars classified as Luminous Blue Variables (LBV's) appear to have the ability
to remove   a large fraction of their outer stellar layers  through violent
ejection processes long before the SN phase is reached.  It is believed that these 
ejections, combined with the effects of the stationary stellar winds,  may determine 
to a large extent the mass of the supernova progenitor (van Marle et al., 2007, 
Smith, 2008, Dwarkadas, 2011). 

The LBV eruptions have been observed since the 1600's, but the mechanisms driving the instability 
have not been identified. Although some scenarios have been suggested 
(Guzik, 2005, Owocki \& van Marle, 2008, Kashi \& Soker, 2010), their confirmation is difficult due 
to the  large distances at which most of these objects lie and  the fact that they are intensely observed 
only after they have undergone an eruptive event. Thus, their pre-eruption characteristics 
are poorly constrained. Furthermore,  observational constraints on their fundamental parameters are
often lacking.

HD 5980 is a multiple system  in the Small Magellanic Cloud  that has been observed 
spectroscopically since the 1960s. It contains an eclipsing binary system plus a third object.
Four well-defined variability timescales are present: 
1) a long-term ($\sim$40 years) variation  (Koenigsberger et al. 2010); 2) sudden eruptive events which were observed in 1993 
and 1994, each lasting less than 1 year ( Barb\'a et al.1995; Koenigsberger et al. 1995, Jones \& 
Sterken, 1997); 3) orbital-phase locked variations with the 19.3-day eclipsing binary period (Breysacher, 
Moffat \& Niemela 1982; Foellmi et al. 2008, and references therein); and 4) ``microvariability" on 
$\sim$30 minute timescale that was observed shortly after the 1993/1994 eruptions  (Sterken \& Breysacher, 1997). 
These timescales disclose the presence of  a variety of physical phenomena in the system.  In particular,
Koenigsberger et al. (2010) have argued that the 1993-1994 eruptions may have been caused by tidal interactions
that became  stronger due to a gradual increase in the stellar radius, on the $\sim$40 year timescale.  
Although the process responsible for the $\sim$40 year timescale is unknown, it is  believed to be the same
as that which occurs in S Doradus -type variables.

The  three luminous objects comprising the HD 5980 system are:  two emission-line 
stars in a close 19.3-day binary  orbit (Breysacher \& Perrier 1980)  and a third 
O-type object (hereafter {\it star C}) whose photospheric absorptions remain relatively stationary on 
the orbital timescale of the binary (Niemela 1988; Koenigsberger et al. 2002).  Following the convention 
introduced  by Barb\'a et al. (1996), we label as {\it Star A} the star ``in front" at orbital phase 
$\phi=$0.00 and {\it Star B} the one ``in front" at the opposite eclipse, which  occurrs at
$\phi=$0.36.  {\it Star A} is the unstable  star of the system (Barb\'a et al. 
1996), and whose spectral type has changed from the Wolf-Rayet subtype WN3 (in 1978-1981; Niemela 1988) to 
WN6 (in 1990; Koenigsberger et al. 1994) culminating in WN11/B1.5Ia$^+$ during the 1994 eruptive 
phase (Drissen et al., 2001; Heydari-Melayari et al., 1997; Koenigsberger et al., 1998a).  {\it Star B} 
is believed to be a WN4 star (Breysacher, Moffat \& Niemela 1982; Niemela 1988). 
Further background on HD 5980's observational characteristics  may be found in  Barb\'a et al. (1996, 1997), 
Moffat et al. (1998), Koenigsberger (2004) and Foellmi et al. (2008).

Although its triple nature implies having to deal with the problem of disentangling the spectra 
of three stars and the possible contribution from a wind-wind interaction region to arrive at the 
eruptor's properties, HD 5980 provides considerable advantages for studying the underlying instability: 
1) its distance and masses can be relatively well-constrained; 2) it is un-obscured by dust; 3) it has 
been widely studied spectroscopically at X-ray, UV and optical wavelengths; and 4) it was well observed  
during the stages prior to the 1993--1994 eruptions  and intensely observed thereafter.  
In this paper we focus on determining the epoch-dependent properties of {\it Star A}.
Section 2 is devoted to a description of the observational 
data; in Section 3  we empirically disentangle the wind velocities of the three stars in the system;
the existence of  correlations between visual magnitude, UV brightness, emission-line strength and 
wind speed are shown to exist in Section 4.  Section 5 contains  the description  of the  
model atmosphere fits to the spectra;  the results are discussed in Section 6 and the conclusions
presented in Section 7.

\section{Observational material}

Ultraviolet spectroscopy  ($\lambda\lambda$1190--2000 \AA) of HD 5980 is available from the 
{\it International Ultraviolet Explorer (IUE)} at low and high resolutions in 1979, 1981,
1989, 1991, and 1994--1995, and at low resolution in 1978 and 1986.  The properties of 
these spectra were analyzed by Moffat et al. (1989) and  Koenigsberger et al. (1994, 1995,
1998a, 1998b), and an overview  of the derived results  may be found by Koenigsberger (2004).
Further UV observations at high resolution were obtained using the {\it Hubble  Space Telescope Imaging
Spectrograph (STIS)} in 1999, 2000, 2002 and 2009.  The properties of these spectra are described
by Koenigsberger et al. (2000, 2001, 2010).  Tables~\ref{tab_uvspec1}  and \ref{tab_uvspec2} list in the first three columns the 
identifying number for each spectrum, the date of acquisition and the corresponding orbital 
phase computed with the Sterken \& Breysacher (1997) initial epoch and orbital period.  For many of 
the  {\it IUE} spectra of HD 5980 a visual magnitude $m_v$  could be derived  (Koenigsberger et al. 1994) 
 from the Fine Error Sensor (FES) counts through the 
FES-$m_v$ calibrations (Perez 1992).  These are listed in column 4 of Tables~\ref{tab_uvspec1}  and \ref{tab_uvspec2}. The FES magnitudes 
have a formal error of 0.07 mag.

The {\it IUE} spectra analyzed in this paper were retrieved from the {\it MAST} data base, which contains 
spectra that have been re-processed with the  Final Calibration.  We find that the difference between the flux
measurements made on these spectra and on the original data products are $<$5\%  for spectra obtained prior 
to 1992, but this difference increases significantly thereafter, in some cases being as high as  
15\%.\footnote{As described by Nichols \& Linsky (1996), there were systematic wavelength-dependent discrepancies
of up to 20\% of the absolute flux calibration in some of the {\it IUE} spectra processed with previous
software.}

Velocities of spectral features were measured using consistent criteria for all  data sets. 
Columns 5--6 of Table~\ref{tab_uvspec1} list the velocities of selected features that will be described in
the next section.  We find that the uncertainty in the speeds measured for the P~Cygni absorption 
components is generally $\sim$100 km/s, which we adopt throughout this paper  unless otherwise noted.
The sources of uncertainty  include the small signal-to-noise ratio of IUE spectra, the contamination  
by lines arising from other atomic transitions and the definition of the continuum level.  In general, 
the new velocity measurements coincide well, within the uncertainties,  with those published previously.

The continuum flux at $\lambda$1850 \AA\ was measured on all UV spectra and is listed in columns 9 and 7
of Tables~\ref{tab_uvspec1}  and \ref{tab_uvspec2}, respectively.   The $\lambda$1850 \AA\ spectral region was chosen to characterize 
the UV continuum level because it is relatively line-free in most of the HD 5980 spectra.  The integrated 
flux of the \ion{N}{4}] 1486 \AA\ and \ion{N}{3} 1750 \AA\  blend was measured and the values are
listed in columns 7-8 of Table~\ref{tab_uvspec1}  and 5-6 of Table~\ref{tab_uvspec2}. Both lines are clearly associated with the
active state of {\it Star A}.  \ion{N}{4}] 1486 \AA\ was first seen in  in 1986, being absent or very
weak previously.  The \ion{N}{3} 1750 \AA\ emission  was first seen  during the declining phase of the 1994
eruption.\footnote{There are unfortunately no UV spectra obtained earlier during this event.}  It  appeared 
as a single emission line in low resolution spectra  but was composed of \ion{N}{3} 2s$^2$2p--2s2p$^2$ and 
\ion{N}{3} 2s2p$^2$--2p$^3$ multiplets. The flux contained in the 
\ion{N}{4}] 1486 \AA\ and \ion{N}{3} emission lines was measured on the low dispersion spectra by 
fitting one or more Gaussian functions.  On high dispersion spectra, the flux was obtained through direct 
integration over the emission feature.  The largest source of uncertainty  resides in the choice 
of the continuum level which  is frequently difficult to define  due to large number of variable emission 
lines  and, on high resolution {\it IUE} spectra, the low signal-to-noise ratio.

The formal uncertainties in the flux calibrations of {\it IUE} and {\it STIS} data  are,  respectively,
$\sim$6\% (Colina, Bohlin \& Castelli 1996) and 1\% (Ma\'{\i}z-Apell\'aniz \& Bohlin 2005).  In general,
the flux levels measured in high and low resolution {\it IUE } spectra differ by no more than the quoted
uncertainty.  However, in the subset of observations obtained in 1994, the high resolution data display 
significantly weaker flux levels than the contemporaneous low resolution spectra.  This  situation might   
be due to tracking problems  during the long exposures. Also, the FES counts are not available during
this time period.\footnote{No report of FES malfunction was found in the IUE Operations Summary,
although a change in the Gyro 5 drift rate is reported to have occurred in October 1994 as well as the
appearance of the DMU anomaly; see http://wise-iue.tau.ac.il/ines/docs/p05.pdf}  Hence, for the analysis 
described in the following sections, we measured the continuum flux only on low resolution IUE data for the 1994 epoch.

An optical spectrum was obtained on 2009 November 7, at phase 0.038 using the Magellan Echellette (MagE) on the  
Clay 6.5m (Magellan-II) telescope on Las Campanas.  We used the 1" slit providing a spectral resolution of 1 \AA. 
Thirteen  echelle orders were extracted covering the wavelength region from 3130 A to 9400 \AA. The 
signal-to-noise ratio ranges from 100 to 200 for a single 150 s exposure. The usual ThAr comparison lamp was 
used for wavelength calibration.  The data were reduced using the special set of IRAF routines available 
for the reduction of MIKE spectra  ({\it mtools} package;
http://www.lco.cl/telescopes-information/magellan/instruments/mike/iraf-tools).
Spectra of the standards Feige 110 and NGC 7293 central star observed during the same night were used to 
derive a sensitivity function. The individual flux calibrated echelle orders were then normalized and 
merged in the final spectrum.

For the modeling discussed below, four representative spectra  corresponding to different observation 
epochs were constructed as follows:

\noindent {\bf Spectrum 1994}: The low resolution {\it IUE} spectra SWP 53218 and LWP 29794 were 
combined with the optical spectrum obtained on the same date, 1994 December 30, at CTIO and 
described in Koenigsberger et al. (1998b). This combined spectrum covers the wavelength region from 
1190 \AA\ to 6900 \AA. The orbital phase is 0.39, and this is the only spectrum not obtained at primary 
eclipse ($\phi$=0) that we analyze.  During this epoch, the spectrum of {\it Star A} was dominant and, 
as in Koenigsberger et al. (1998b), the contribution from {\it Star B} is  neglected.
 
\noindent{\bf Spectrum 2000}: The {\it HST/STIS} spectrum obtained on 2000 April 20 at phase 0.00. This 
spectrum covers the spectral range 1150 to 10800 \AA. 

\noindent{\bf Spectrum 2002}: The {\it HST/STIS} observation of 2002 April 4, obtained  at phase 0.99 was 
combined with the  {\it FUSE} spectrum P2230101 taken on 2002 July 27 at phase 0.00.  The systematic 
long-term variations in HD 5980 were relatively small between 2002 and 2009, justifying that we combine 
these two spectra obtained $\sim$3 months apart. 

\noindent{\bf Spectrum 2009}:  The {\it HST/STIS } spectrum secured on 2009 September 9 at $\phi=$ 0.99  was 
combined with the Magellan-II optical spectrum obtained on 2009 November 7 at phase 0.04. As in the previous case, 
the long-term variability is not expected to be significant.  Of greater concern is the fact that the optical
spectrum was obtained 0.05 in phase later than the UV spectrum.  At $\phi$=0.04 the eclipse of {\it Star B}'s
continuum-emitting region is partial (see light curve in Foellmi et al. 2008). 

\section{Disentangling the wind velocities \label{sec_vel}}

The terminal wind speed $V_\infty$ is generally derived from the saturated portion of the
P~Cygni absorption component; i.e., where the intensity reaches zero level, generally
referred to as $V_{black}$ (Prinja et al. 1990).  In the case of a binary system, the
maximum extent of the saturated  profile corresponds to the speed of the slower wind
in the system.  The faster wind also has a $V_{black}$, but the minimum
intensity level lies at  the continuum level of the star whose wind is slower.  Thus, as shown in
Georgiev \& Koenigsberger (2004), the absorption profile  presents a step-like appearance.  
We will refer to the second $V_{black}$ as $V_{plateau}$.
For single Wolf-Rayet stars, it is generally  found  that for saturated lines $V_{black}\sim$0.70 $V_{edge}$
(Prinja et al. 1990; Eenens \& Williams 1994), where $V_{edge}$ is  the location where the P~Cygni
absorption  meets  the continuum level. The observation that $V_{edge}>V_{black}$ is generally
attributed to an unspecified type of ``turbulence" or to non-monotonicity in the wind.  In the 
case of optically thin lines, the P~Cygni absorptions do
not reach the zero flux level. In this case,   $V_{edge}$ provides information on  the
maximum wind speed attained within the particular line-forming region.

Interpreting the P~Cygni absorption components in HD 5980 is  difficult  because  3 massive 
and hot stars contribute to its spectrum.  Koenigsberger et al. (1998a) adopted the purely 
empirical approach of measuring $V_{black}$ and $V_{edge}$ in {\it IUE} spectra obtained over 
the years 1979--1995.  They found a persistent  component at $V_{black}  \sim -$1740 km/s indicating 
the presence of a stable wind with this speed  and  showed that the wind speed of {\it Star A}  
had undergone changes from  $<$500 km/s to $\sim$ 1600 km/s during and after the 1994 eruption.  
However, although they suspected that  {\it Star A}'s wind speed had  been as high as  $\sim$3000 km/s 
in 1979, disentangling its contribution to the P~Cygni absorptions from that of the other two stars 
proved very challenging.   

The more recent data have now clarified the picture, partly
because of its greatly improved quality and partly because it has been possible to obtain
UV observations during orbital phase $\phi$=0.0, when {\it Star A} occults {\it Star B}. 
The size of the eclipsing disk of { \it Star A} is  a factor of 1.5 larger than {\it Star B} at  
minimum  brightness (Perrier et al. 2009) and significantly larger during its more active state.
It is therefore valid to assume that  at orbital phases $\phi\sim$0.0 {\it Star A} eclipses  {\it star B}'s     
continuum-forming disk as well as the  P Cyg absorption forming region of {\it star B}'s wind.  Thus,
the P~Cygni absorptions at $\phi\sim$0.0 provide the wind speeds of {\it Star A} and {\it Star C}, without
contamination from {\it Star B}.

\subsection{{\it Star C}: the ``third" object \label{sec_starC}}

\begin{figure}
\plottwo{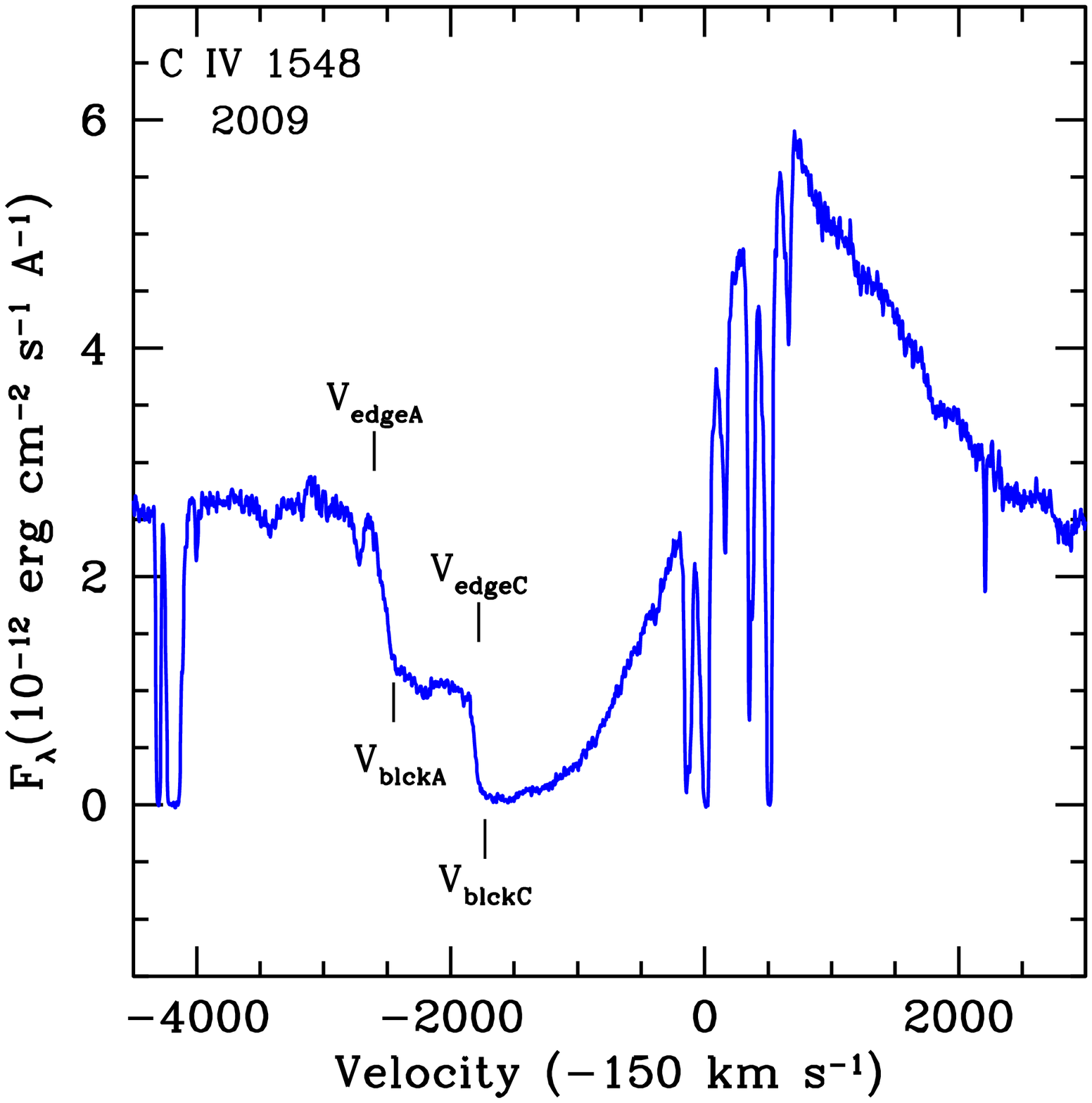}{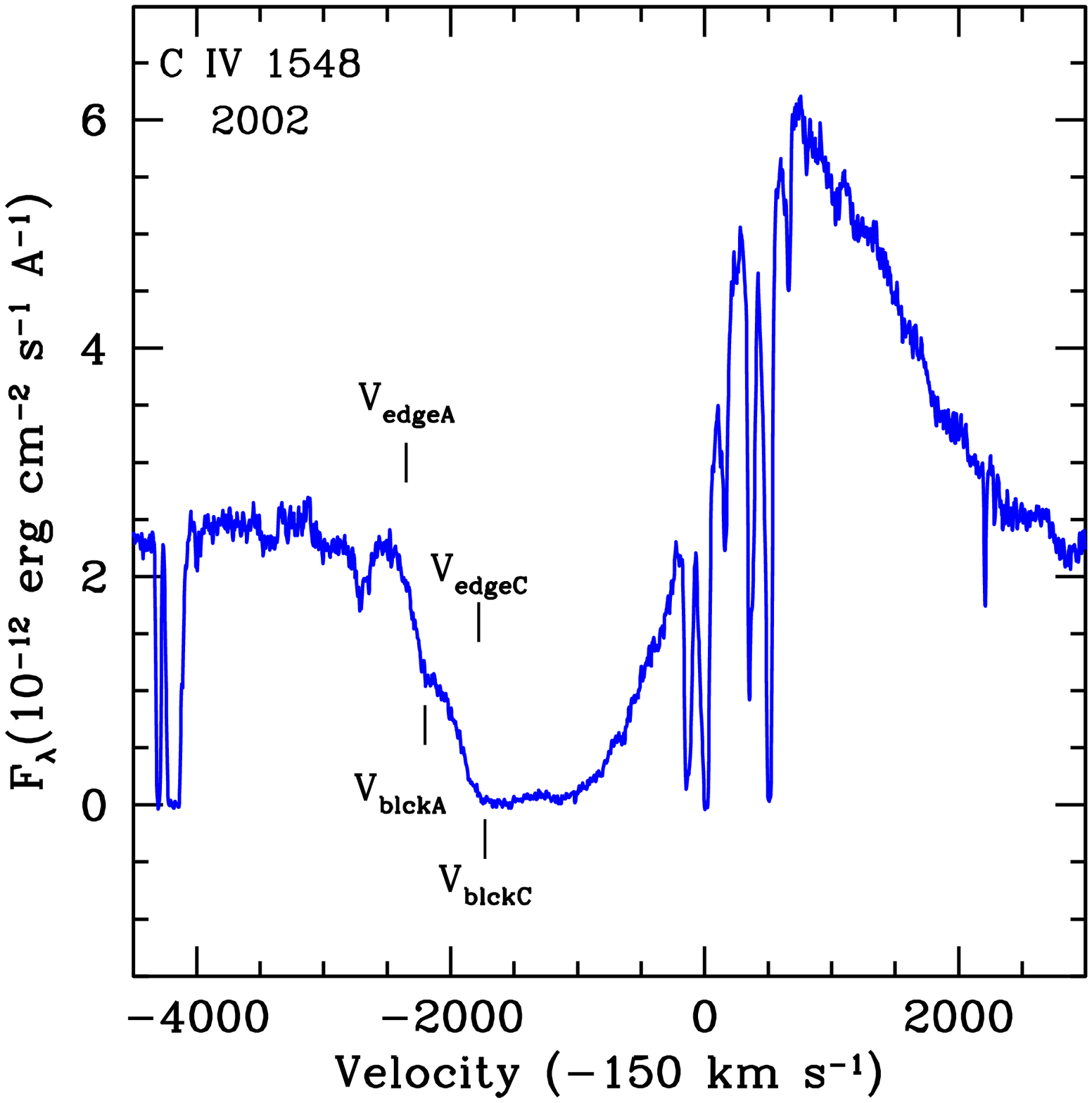}
\caption{The C IV 1548-1550 P~Cygni doublet observed in 2009 (left) and 2002 (right) on a velocity scale 
centered on the 1548 \AA\ laboratory wavelength corrected for the SMC motion ($+$150 km/s).
\label{fig_civ1}}
\end{figure}

Fig.~\ref{fig_civ1} is a plot of the \ion{C}{4} 1548/1550 \AA\ doublet observed in the 
{\it HST/SITS} spectra of 2009 (left panel) and 2002 (right panel) at orbital phase 0.99. 
A clear ``step" is observed in  the 2009 spectrum, providing two velocity values:  $V_{black}$= $-$1760 km/s 
and $V_{plateau}$=$-$2440 km/s.\footnote{Note that the separation between these two $V_\infty$ values  is $\sim$700 km/s, 
significantly greater than the separation between the two \ion{C}{4} doublets (500 km/s).}  The 
``step" is not as pronounced  in the spectrum of  2002 (Fig.~\ref{fig_civ1}, right), but still
two velocity values may be measured:$V_{black}$= $-$1780 and $V_{plateau}$=$-$2210 km/s. 
$V_{black}$= $-$1770$\pm$10 km/s is consistent with the persistent wind velocity measured by
Koenigsberger et al. (1998a) for all epochs prior to 1996.  Since only two stars are visible at $\phi$=0.99,
it is now clear that this stable component is formed in {\it Star C}.  These spectra also give us the 
wind speed of {\it Star A} in 2002 and 2009, a topic to which we shall return  in Section 3.3.

\subsection{{\it Star B}: the elusive companion \label{sec_starB}}

Of the three distinguishable objects in  HD 5980, {\it Star B} is the most elusive.  From the 19.3d
radial velocity variations and the eclipse light curve, there is no question that {\it Star B} is a hot 
and massive object.  However, given that {\it Star A} is such a prominent source of emission-lines, it is
difficult to determine the fraction of the emission-line spectrum that may be attributed to the
wind of {\it Star B}.  On the other hand, during the eclipse when {\it Star B} is ``in front", the
wind region of {\it Star A} where the fast portions of its P~Cygni absorption components are formed is
occulted.  This eclipse occurs at $\phi$=0.36 and can be used to estimate {\it Star B}'s wind speed.

\begin{figure}
\plottwo{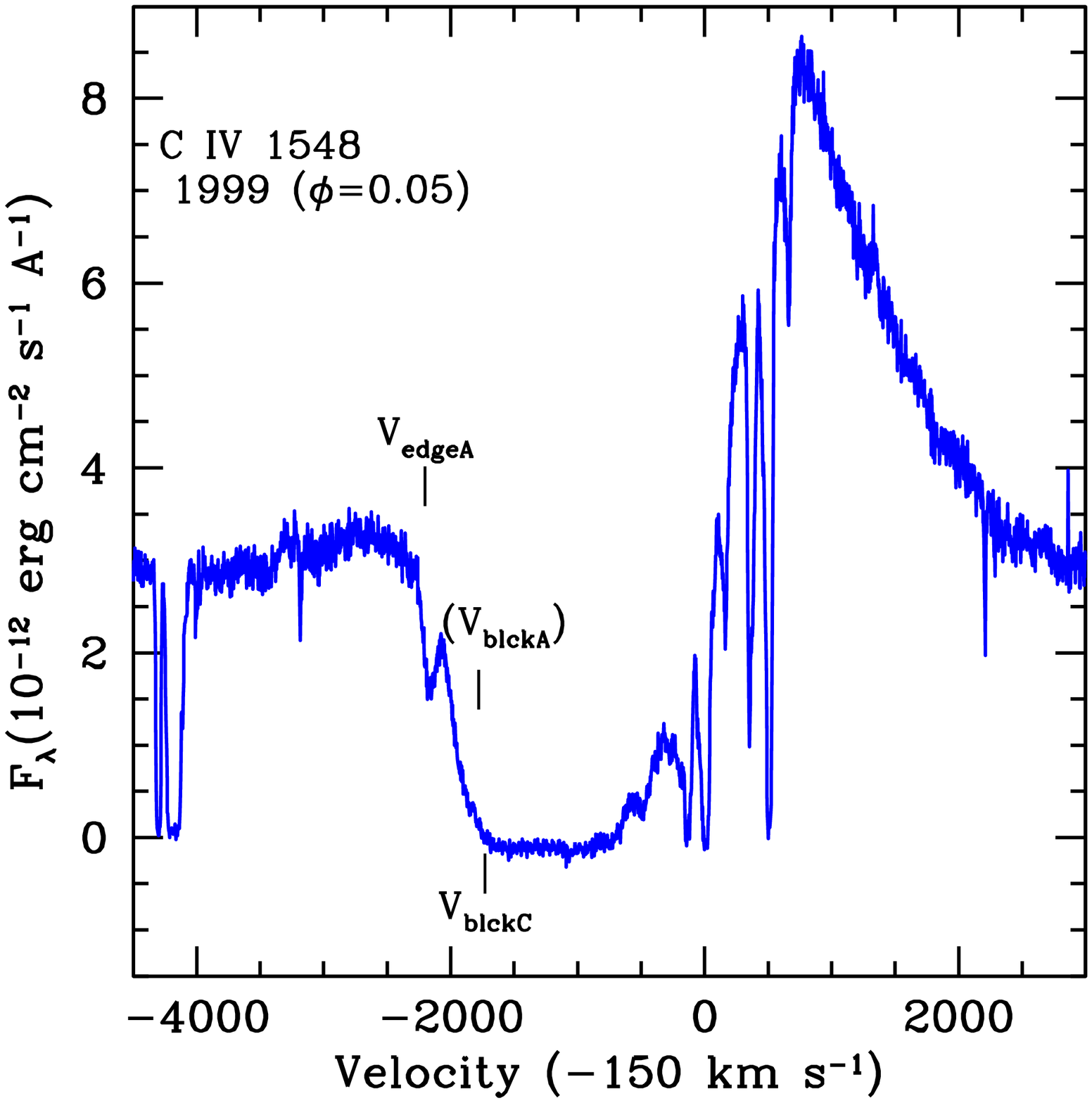}{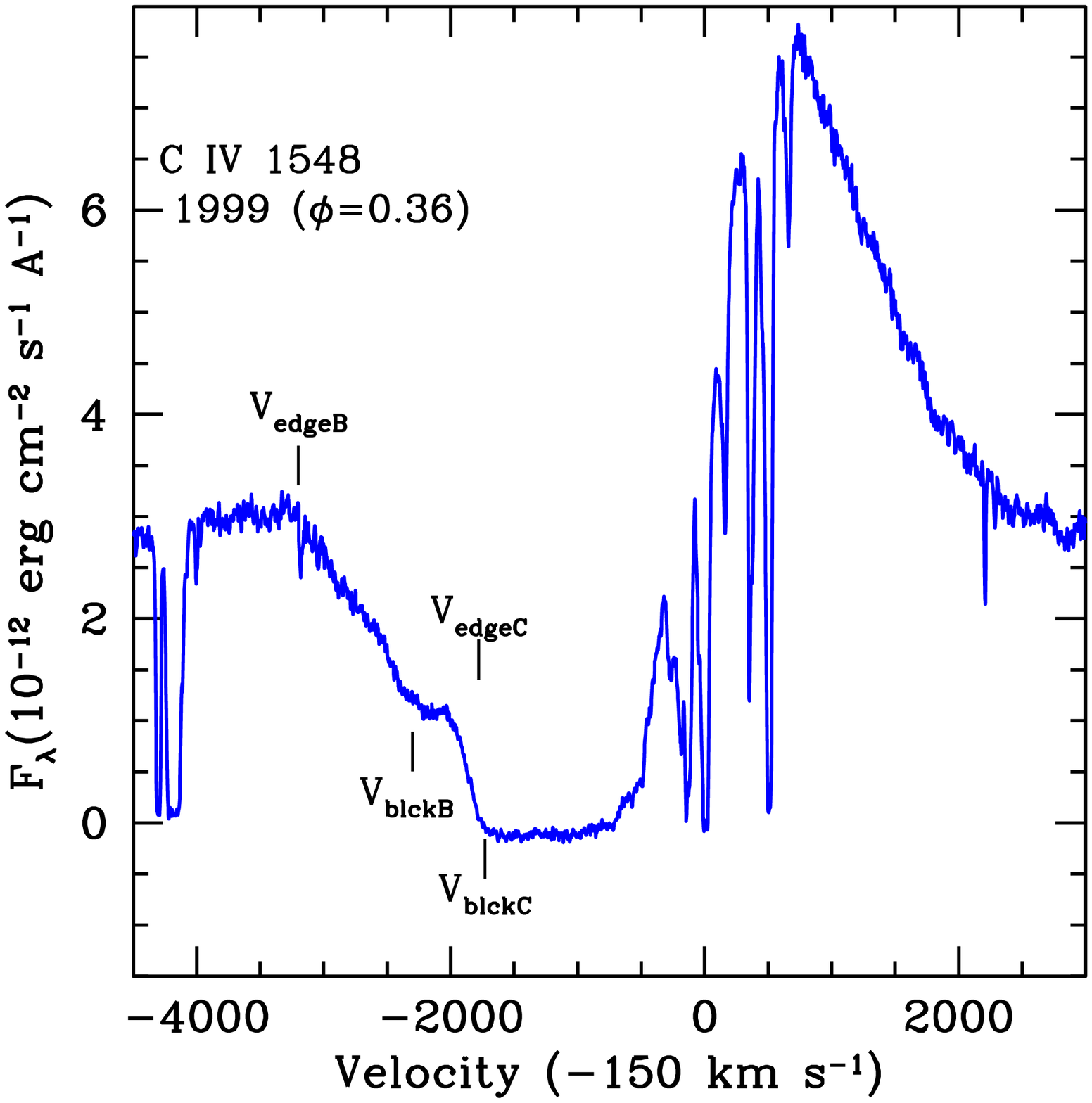}
\caption{The  \ion{C}{4} 1548-1550 \AA\  P~Cygni doublet observed in 1999 at orbital phases 0.05 (left) and
0.36 (right), the latter corresponding to the eclipse when {\it Star B} is ``in front".  The velocity scale
is centered on the 1548 \AA\ laboratory wavelength corrected for the SMC motion ($+$150 km/s). The
large difference of 950 km/s between $V_{edgeB}$ and $V_{blckB}$ in right panel indicates a large amount of
``turbulence".
\label{fig_civ2}}
\end{figure}

In 1999,  5 {\it HST/STIS} spectra of HD 5980 were obtained over a single orbital cycle (Koenigsberger et al. 2000), 
including a spectrum at $\phi$=0.36.  The line-profile of \ion{C}{4} 1548-50 \AA\ at this phase  is shown
in Fig.~\ref{fig_civ2} (right) where we observe $V_{black}=-$1720 km/s, associated with {\it Star C} as discussed
in the previous section.  In addition, there is a short plateau that ends at $V_{plateau}$=$-$2300$\pm$100 km/s,
followed by  a very extended slope over which the line rises towards the continuum level extending 
to V$_{edge}$= 3100 km/s.  We associate the $V_{plateau}$ with the terminal wind speed of {\it Star B}.
The value of V$_{edge}$ is consistent with the relation  $V_{black}/V_{edge}\sim$0.76 typically observed 
for WR stars (Prinja et al. 1990).
It is important to note that the eclipse at $\phi$=0.36 is not total, since {\it Star B}'s radius is
smaller than that of {\it Star A}.  However, the wind velocity component along the line-of-sight from
the unocculted portion of {\it Star A} is rather slow and any P~Cygni absorption formed in this star lies
close to the rest-frame velocity.  Thus, the very large value of V$_{edge}$ is most likely associated
with ``turbulence" in  the wind of {\it Star B}.

\begin{figure}
\plotone{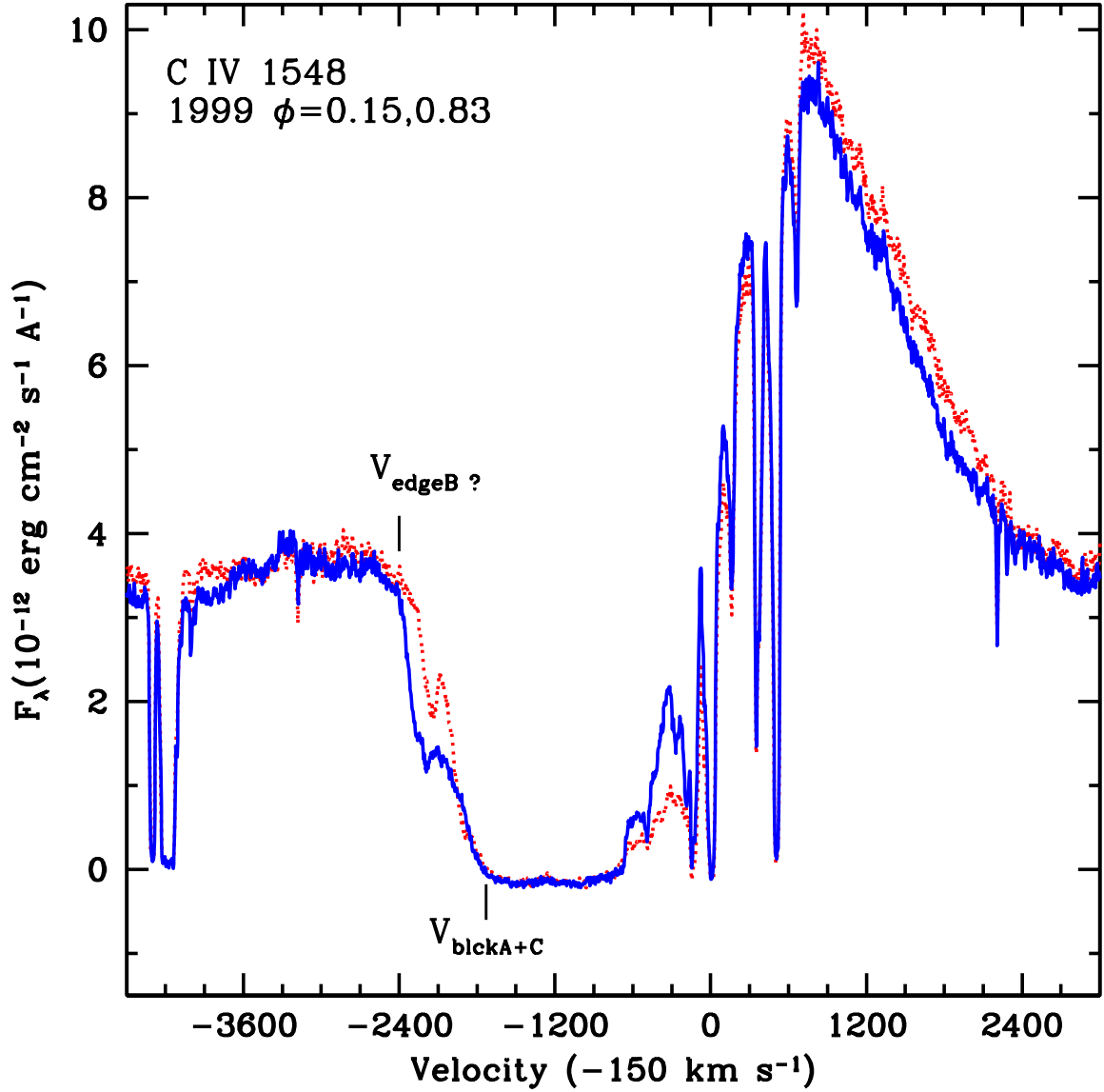}
\caption{The C IV 1548/1550 P~Cygni doublet observed in 1999 at orbital phases 0.15 (dotted line) and
0.83. These phases correspond to the largest projected orbital motion of the stars in the 19.3d binary. 
The ``turbulet" edge velocity of $-$3100 km/s seen at $\phi$=0.36 is absent, possibly due to the wind-wind
interaction region.  The velocity scale is centered on the 1548 \AA\
laboratory wavelength corrected for the SMC motion (+150 km/s).
\label{fig_civ3}}
\end{figure}

At other orbital phases the picture is not as clear, however.  Two observations at elongations were
obtained as part of  the same 1999 set of {\it HST/STIS} spectra mentioned above.  The elongation
phases observed are $\phi$=0.15 and $\phi$=0.83, when {\it Star B} is approaching the observer  and receding,
respectively.  At $\phi$=0.15 both the $-$2300 km/s plateau and $-$3100 km/s extended absorption should
be clearly visible.  However, the spectrum shows only $V_{edge}\sim-$2400 km/s, as illustrated in 
Fig.~\ref{fig_civ3}.  This suggests that the ``turbulent" component is suppressed in the regions that are 
viewed expanding along-the-line of sight to the observer at $\phi$=0.15.  This may be a consequence of 
the presence of the wind-wind interaction region, an issue that will be addressed in a forthcoming paper.
For the present investigation, it suffices to  keep in mind that {\it Star B}'s wind velocity is 
V$_B \sim$ 2300 km/s and that its contribution to the system's spectrum is important anly when it 
is ``in front" of {\it Star A}. 

\subsection{{\it Star A}: the eruptor \label{sec_starA}}
 
In Fig.~\ref{fig_civ2} (left) we illustrate the spectrum obtained in 1999 at $\phi=$0.05. This phase  
is close to primary eclipse,  and {\it Star A} is between us and {\it Star B}. Here there is no
plateau in the absorption line-profile indicating that {\it Star A}'s terminal wind velocity is 
similar to that of {\it Star C} and both stars contribute to $V_{black}$=$-$1720 km/s.  In Section 3.1 
we determined the wind speed of {\it Star A} in 2002 and 2009 to be 2210 and 2440 km/s, respectively.
This  trend for increasing wind speeds between 1999 and 2009 is one that can be traced back to 
late 1994, at which time speeds $<$500 km/s were recorded (Koenigsberger et al. 1998a). This is the inverse
of HD 5980's behavior in the epochs preceding  the eruption;   between 1979 and 1991 wind speeds steadily 
declined.

In the  1979 {\it IUE} spectrum obtained at $\phi=$0.91 a plateau extending to V$_{plateau}=-$2670 km/s  is 
clearly present in the C IV 1550 P~Cygni absorption, followed by a gradual rise reaching  the continuum 
level at $-$3200 km/s.  The other spectra obtained in 1979 also display the plateau  which, for example, 
in the case of the $\phi=$0.48 spectrum  extends to $-$2750 km/s.   This is similar to the $\phi$=0.36 spectrum 
discussed in Section 3.2. But because  none of the 1979-1980 spectra were obtained during eclipse, it is not 
possible {\it a priori} to determine which of the two stars is responsible for the different fast components.  
However, since the speed determined from spectra obtained in other epochs when {\it Star B}  eclipses {\it Star A} 
are $\sim-$2300 km/s, we are led to conclude that the $\sim-$2700 km/s component in 1979 corresponds to the wind 
of {\it Star A}.  A similar analysis leads to the conclusion that by 1991 {\it Star A}'s wind velocity 
decreased to $\sim -$2200 km/s.

The individual contribution from {\it Star A} and {\it Star B} may also be identified in other P~Cygni lines
such as He II 1640 and N V 1240, lines to which the contribution from {\it Star C} is negligible.   
The N V line displays a flat portion, analogous to  $V_{black}$, but which does not reach the zero 
intensity level. Its extent provides the velocity of the slower wind, either that of {\it Star A} or
{\it Star B}, depending on the epoch.  A second plateau provides the velocity of the faster wind. 
The measurements of these two plateaus are listed in column 6 of Table 1. 

The He II 1640 line arises from a transition between two excited states, and thus, the strength of 
its P~Cygni absorption is weaker than that of the C IV and N V resonance transitions.  Here, the simplest 
measurement to perform is that of V$_{edge}$. Because there is no ``black" portion in the absortion, it is 
not clear to what extent ``turbulence" contributes to  the edge velocity.  In addition, some of the spectra 
show a break in the absorption profile, similar to the plateau seen in the resonance lines, and which 
may be associated with the second star in the system. The measured values of V$_{edge}$ and of the 
plateau (when visible) are listed in column 5 of Table 1. We find that these values are consistent with 
the velocity for {\it Star A} derived from C IV and N V. Because the contribution from {\it Star C} 
may be neglected in N V 1240 and He II 1640, we use these lines for the analysis that will be presented 
in the next section.

\begin{figure}
\plotone{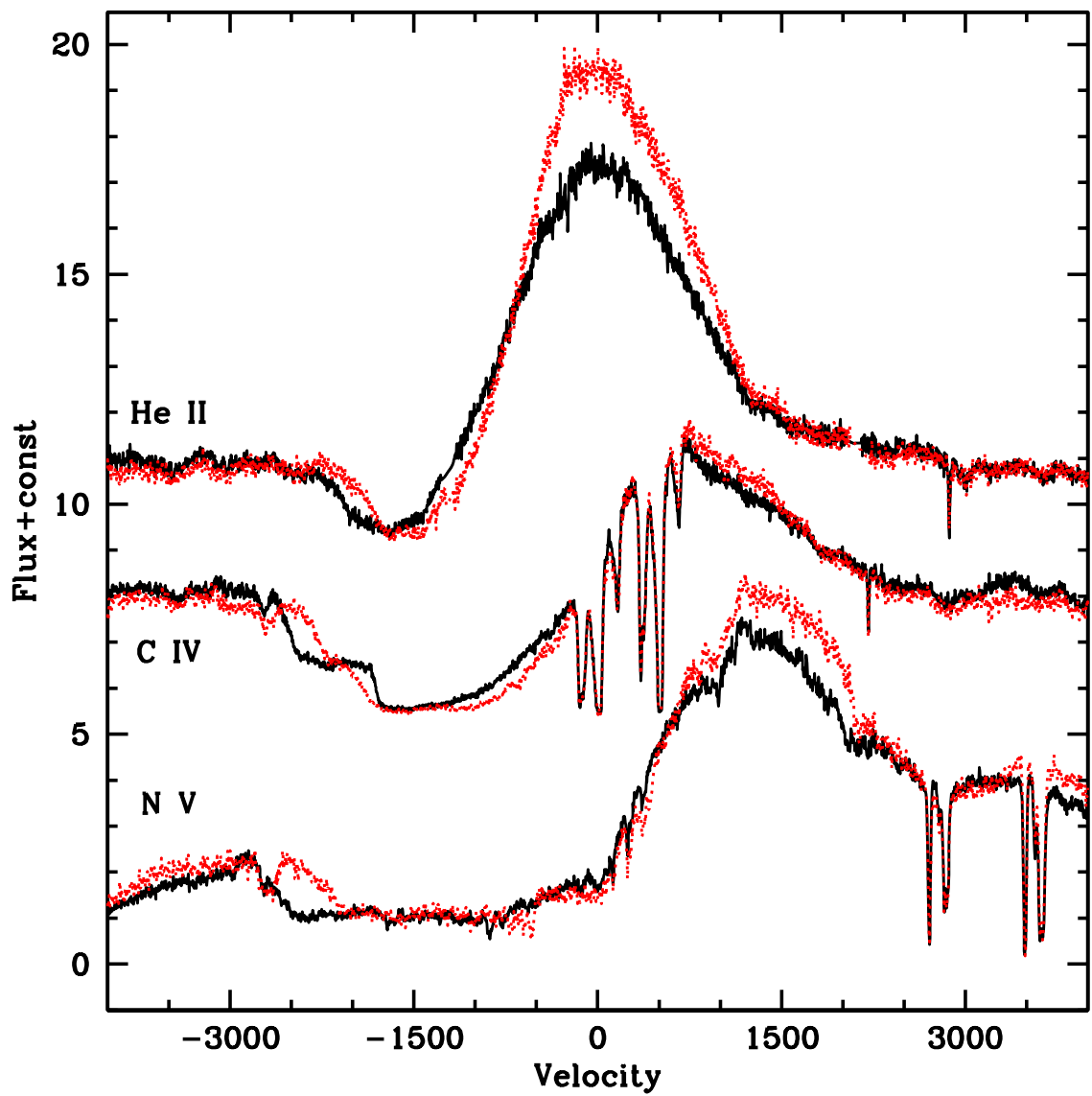}
\caption{Montage of the N V, C IV and He II P~Cygni profiles in the {\it HST/STIS} spectra
obtained in 2002 (dots) and 2009. Both spectra correspond to the same orbital phase
$\phi=$0.99.  The most striking differences are the strength of the emission line and the 
extent of the absorption component.  Note the ``step"-like shape of the C IV absorption 
component in 2009 due to the contribution from {\it Star C} (-1760 km/s) and {\it Star A} 
(-2440 km/s).  Velocity scale is centered on the laboratory wavelengths of NV1238.821, 
C IV 1548.187 and He II 1640.47, with a correction of $-$150 km/s to account for the relative 
SMC motion.
\label{fig_civ5} }
\end{figure}

The behavior of the 3 lines is illustrated in Fig.~\ref{fig_civ5} for the epochs 2002 and 2009, showing
that they all  display the wind velocity increase  that took place between 2002 and 2009.

A summary of our estimated wind speeds for the three stars at different epochs is given in
Table~\ref{tab_star_velocities}.  These speeds, as those of Table~\ref{tab_uvspec1}, are corrected for the 
systematic velocity of the SMC ($+$150 km/s) but do not include a correction for orbital motion.  
It is important to note that, except for the 1979-1980 epochs, the He II 1640 values of $V_{edge}$ are 
systematically smaller than the values derived from C IV which are listed in Table~\ref{tab_star_velocities}.  
It is no clear whether this is caused by the presence of additional emission lines that ``fill in" the 
absorption near its edge\footnote{The numerous absorptions associated with {\it Star A}'s active state 
are absent in 1979-1980} or whether this is a consequence of the excitation structure in the wind.

\section{Empirical correlations}

\begin{figure}
\plotone{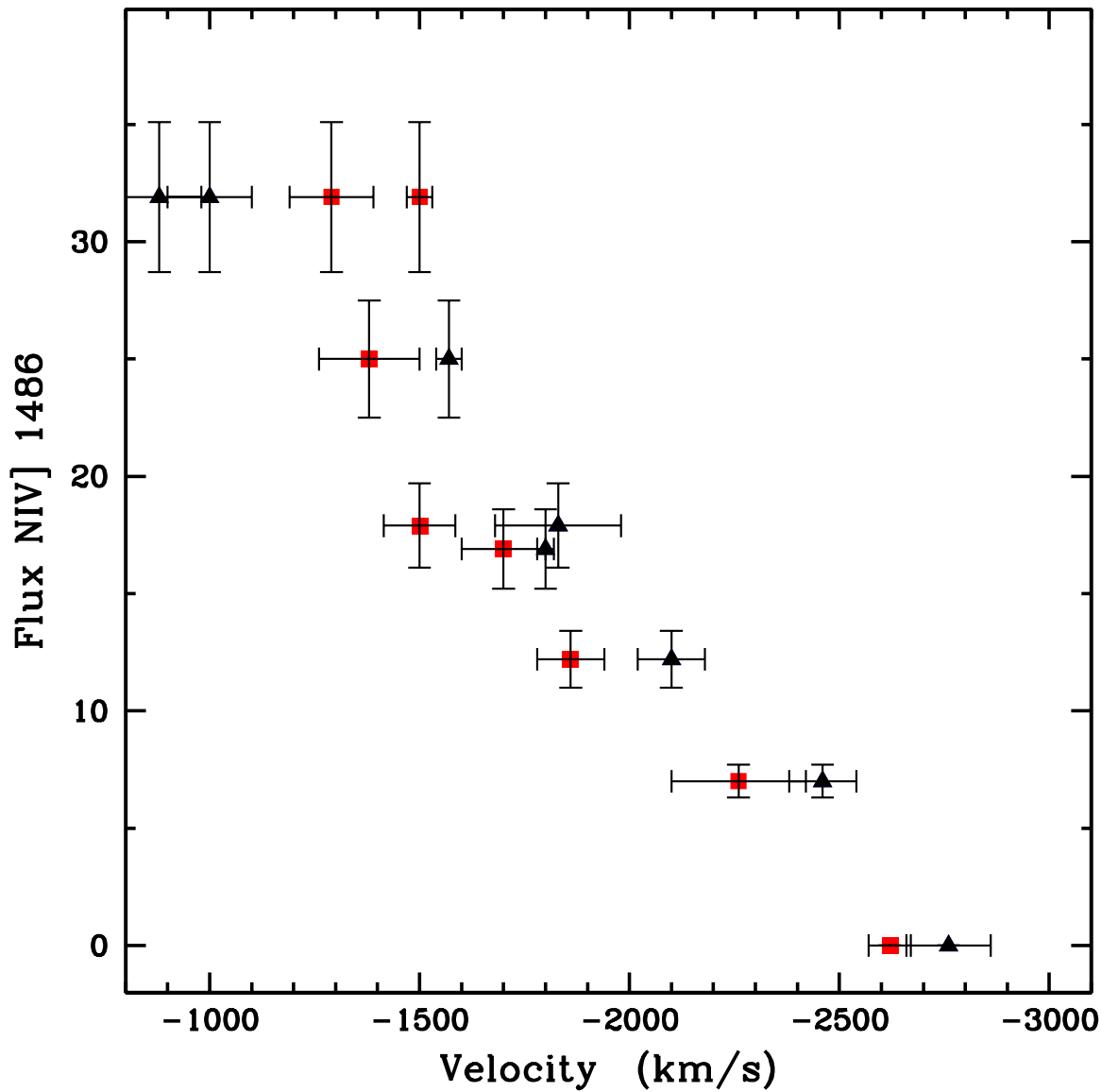}
\caption{ Line strength {\it vs.} wind velocity correlation. The N IV] 1486 \AA\  emission-line flux   
plotted against the wind velocity derived from P~Cygni absorption profiles. Only data from spectra 
obtained around $\phi=$0.00 are plotted. Squares
correpond to $V_{edge}$ of He II 1640 \AA\  and triangles correspond tothe saturation portion of 
 N V 1238. Error bars in velocity correspond to estimated uncertainties in each individual measurement
while the flux uncertainty is estimated at 10\%.
\label{fig_niv_vel}}
\end{figure}

We have previously shown (Koenigsberger et al. 2010) that the emitted flux in the UV and optical lines  
increased during the pre-eruption epochs and decreased after $\sim$1999. Changes in the continuum flux 
levels have also occurred.  Here we show the existence of 2  correlations and their corollaries:

\noindent {\bf Line strength {\it vs.} wind velocity}:  A decrease in the wind velocity is accompanied by an 
increase in the strength of the emission lines; i.e., $F_{lines} \sim V_{wind}^{-1}$. Fig.~\ref{fig_niv_vel} 
shows the flux contained in  the NIV] 1468 emission line  plotted against the P~Cygni absorption-line 
velocity of \ion{He}{2} 1640 \AA\ and \ion{N}{5} 1240 \AA. Only data from spectra obtained around $\phi=$0.00 
are plotted, so the velocity clearly corresponds to that of {\it Star A}.  

\begin{figure}
\plotone{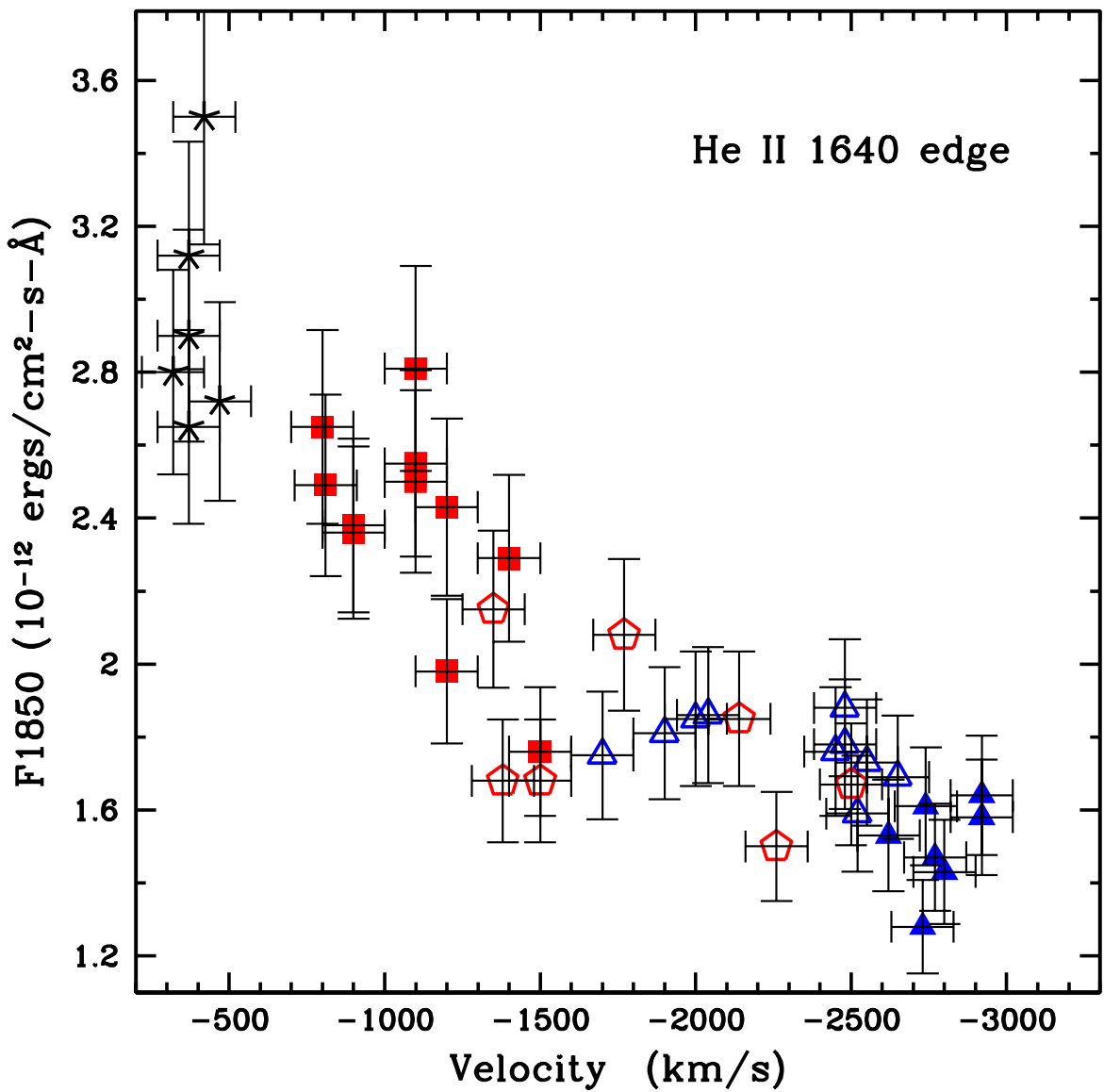}
\caption{Continuum intensity {\it vs.} wind velocity correlation. The UV continuum flux level, F$_{1850}$, plotted
against the wind velocity derived from $V_{edge}$ of He II 1640. 
Symbols indicate:  filled triangles -- 1979--1981; open squares -- 1986; open triangles -- 1989--1991;
stars -- 1994;  filled squares -- 1995; and pentagons -- 1999-2009. For 1994, we use values of
F$_{1850}$ measured on the low dispersion spectra.
\label{fig_f1850} }
\end{figure}

\noindent {\bf Continuum intensity {\bf vs.} wind velocity}: A decrease in the wind velocity is accompanied by an 
increase in the strength of the continuum in the UV ($\lambda$1850 \AA) and in the visual range; i.e., 
$F_{contin} \sim V_{wind}^{-1}$.  This is illustrated in  Fig.~\ref{fig_f1850}, where the UV continuum flux
is plotted against the wind velocity measured from $V_{edge}$ of He II 1640 \AA.

\begin{figure}
\plotone{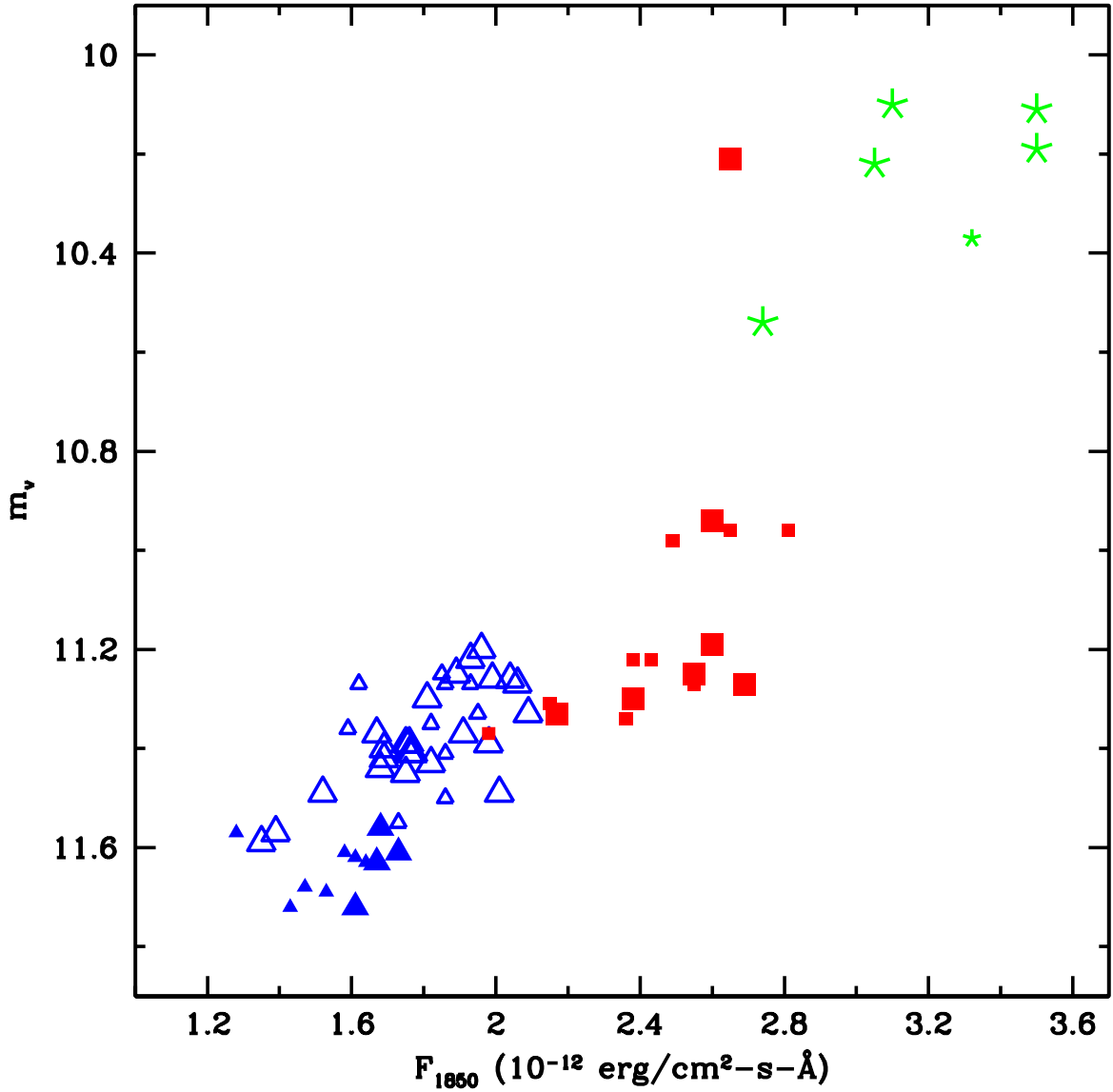}
\caption{Visual magnitudes derived from the {\it IUE} FES counts and the
continuum absolute flux values at 1850 \AA. Symbols indicate: filled triangles -- 1979-1981;
open triangles -- 1989-1991; stars -- 1994; filled -- 1995; Large/small symbols indicate
low/high resolution spectra.  The increasing trend in both wavelength regions suggests an increase
in bolometric brightness.
\label{fes_f1850_calibration2}}
\end{figure}

These two correlations indicate that changes in the wind velocity are accompanied by changes in  
the continuum forming region as well as  in the more extended line-forming regions. That is, the entire wind structure
is affected.  It is also important to note that the UV  and the visual continuum levels increase or decrease 
together,  as illustrated in Fig.~\ref{fes_f1850_calibration2}.   Hence, the phenomenon is not simply due to a 
redistribution in wavelength of the continuum energy.

\begin{figure}
\plottwo{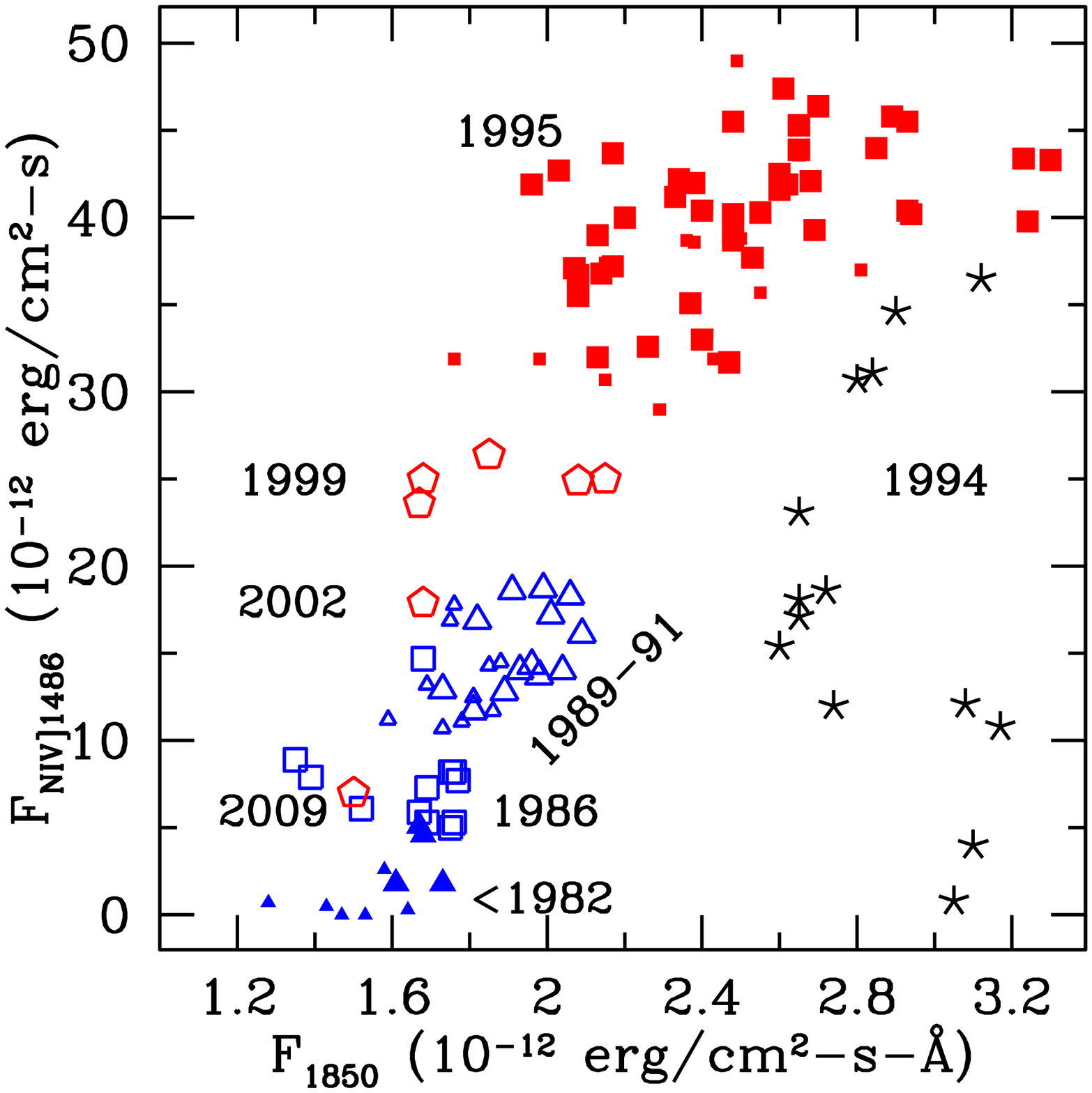}{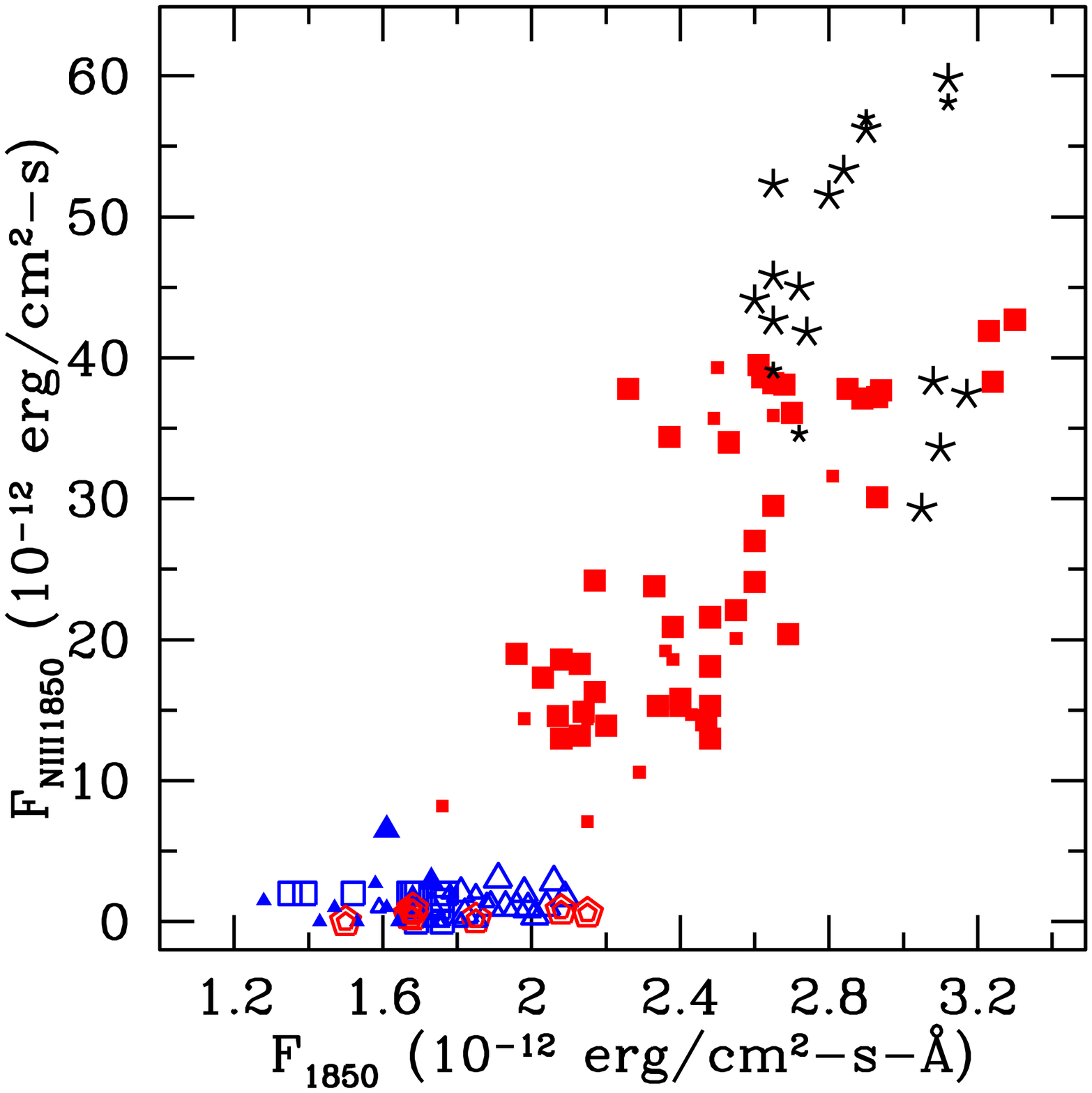}
\caption{Absolute flux contained in the N IV]1486 \AA\ (left) and N III 1750+1754 \AA\ (right)
emission lines plotted as a function of the absolute continuum flux at 1850 \AA.  The different 
epochs are indicated by different symbols:
filled triangles -- 1979--1981; open squares -- 1986; open triangles -- 1989--1991; stars -- 1994;
filled squares -- 1995; and pentagons -- 1999-2009. Large symbols correspond to values obtained from
the IUE low resolution spectra and HST/STIS data; small symbols to high resolution IUE spectra.
Uncertainties are $\sim$10\% of the flux values.
\label{fig_niv_cont}}
\end{figure}

A corollary of the above relations is that the continuum brightness and emission line strengths increase or 
decrease together;  $F_{cont} \sim F_{lines}$.  This is illustrated in  Fig.~\ref{fig_niv_cont},
where the strengths of N IV] 1486 and N III 1750  are plotted against the flux in the continuum  at 1850 \AA.  
The different epochs of observation lie in different locations within this diagram.  These results suggest 
that the physical phenomenon causing the changes in {\it Star A} involves an overall increase or decrease 
in the energy that is emitted; i.e., bolometric luminosity variations.  

\begin{figure}
\plotone{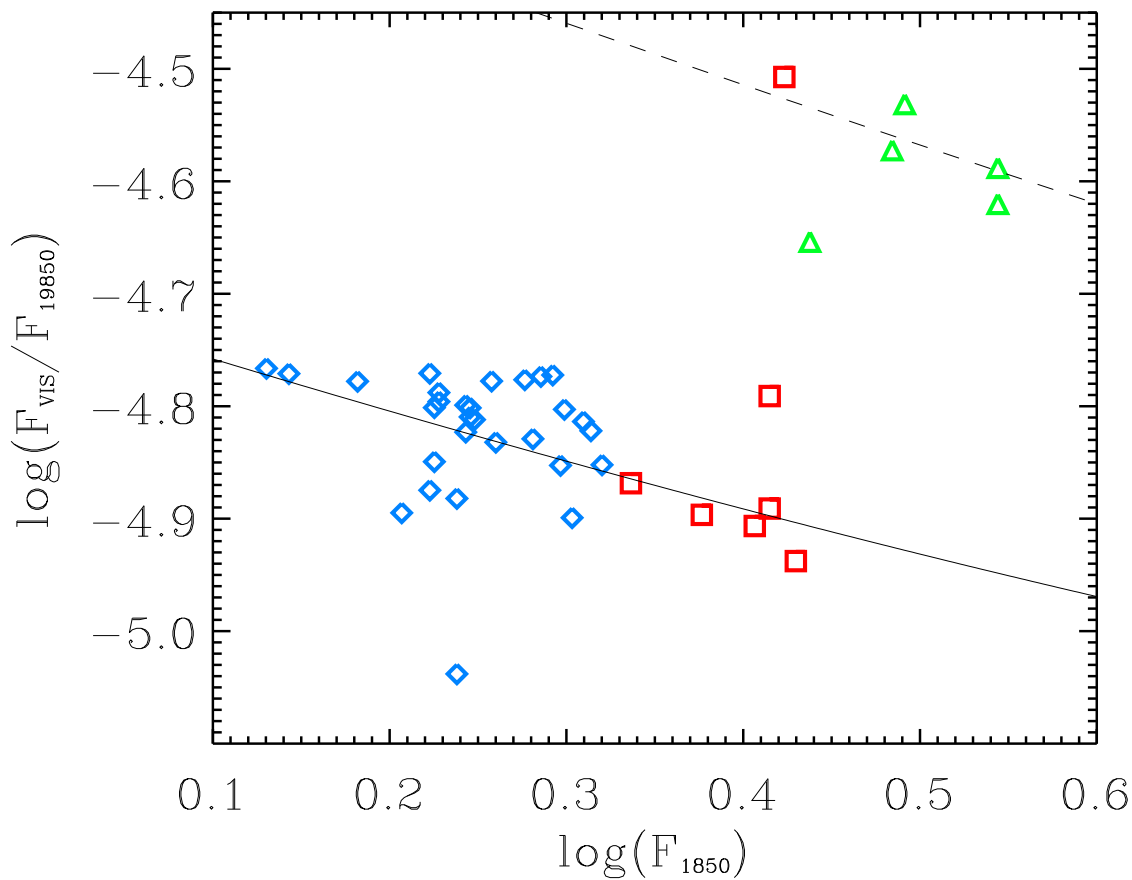}
\caption{Ratio of the visual flux, $F_{VIS}$ converted from the {\it IUE} FES magnitudes and 
$F_{1850}$ from Table 2, plotted against $F_{1850}$. Small open squares - data before 1994;
triangles - data of 1994; large squares - data after Jan 1995.
The continuos line corresponds to the relation of $F_{VIS}$/$F_{1850}$ {\it vs} $F_{1850}$ for 
a black body energy distribution flux of different temperatures  assuming a constant luminosity. 
The dashed line is the same relation but with a luminosity that is 6 times larger, implying that
the 1994 eruption involved a significant change in the luminosity.
\label{fig_fes_cont}}
\end{figure}

Let us momentarily assume that {\it Star A}'s  continuum emits as a black body.
The ratio F$_{vis}$/F$_{1850}$ = $10^{-0.4*m(FES)}/F_{1850}$ at constant luminosity  is an almost linear function of the flux at 1850 \AA\ with 
F$_{vis}$/F$_{1850}$ decreasing with the increasing temperature (continuos line on Fig.~\ref{fig_fes_cont}).
The observed points in Fig.~\ref{fig_fes_cont} are clearly separated in two groups. The points 
before the eruption and after 1995 lie along a  curve similar to the one described by the black body, while the points 
obtained during the eruption are displaced from this correlation.  The temperature decreased during the eruption  
but contrary to what might be expected, the absolute flux at 1850\AA\  increased, as seen in the observed 
F$_{vis}$/F$_{1850}$ correlation. To account for this, the luminosity must have increased by a factor of 
$\sim$6 (dashed line on Fig.~\ref{fig_fes_cont}). This leads to the conclusion that the 1994 eruption 
involved a luminosity increase, a conclusion that is strengthened by the results of the CMFGEN modeing
described in the next sections.  The dispersion  of the points around the black body curve during the other 
epochs points to some further changes in the luminosity although  at a much smaller scale. 

The interpretation of  the  line strength {\it vs.} wind velocity correlation is not straightforward. 
The larger V$_\infty$ increases the transformed radius,    

\begin{equation}
\label{eq_rt}
R_t = R_{star} \left(\frac{V_\infty/2500 \, km/s}{\dot M/10^{-4} \, M_\odot/yr}\right)^{\frac{2}{3}}
\end{equation}

\noindent (Schmutz et al. 1989) reducing the line strength.  Thus, having in mind that V$_\infty$ has been
changing,  one cannot directly interpret the weaker lines as a result of a lower mass loss rate.  
To proceed further in the interpretation of the line intensity variations it is necessary to model the 
spectra of {\it Star A}, which will be done in the next section. 

\section{CMFGEN models of {\it Star A} and {\it Star C}}

As previously described, {\it Star A} undergoes two  modes  of large-scale variability:  1) A long term 
($\sim$40 years) S Dor type variability and 2) An eruptive mode, as occurred in 1993-1994, shortly before the 
maximum of the S Dor  cycle.  The observed changes in lines and continuum fluxes that were described in 
Section 4  point to two separate physical states of the wind corresponding  to these two modes. In order to gain 
further insight into the processes which drive the variability,  we modeled the  spectra  obtained in 1994, 
2000, 2002 and 2009 which are described in Section 2.  These spectra correspond
to  times when the emission from {\it Star A} dominates over {\it Star B}'s emission.  However, {\it Star C} is 
always present in the observations, so a model for this star was also constructed in order to adequately compare 
the models with the data.  All comparisons are made against  the sum of the fluxes 
from the {\it Star A} and {\it Star C} models.  

We modeled the spectrum of {\it Star A} using a new version of CMFGEN code. In this version the spectrum forming 
region of the star is modeled as a hydrostatic photosphere and a wind attached to it. The wind is described with the usual 
$\beta$ law. The photosphere is specified with its gravity ($\log g$) and temperature T$_{10}$ at the Rosseland 
optical depth $\tau_{Ross}$ = 10. This large value of $\tau_{Ross}$  is chosen  so that models with different 
mass loss rates are comparable. The density of the  photosphere is calculated to satisfy the equation 

\begin{equation}
\bigtriangledown P = -\rho(g_{grav} - g_{rad})
\end{equation}

\noindent where  g$_{grav}$ is the specified gravitational acceleration and g$_{rad}$ is the calculated acceleration 
due to the radiation pressure. The solution defines the density as a function of the radius. This defined density and 
the adopted  $\dot M$ together with  
the equation of continuity defines the velocity as function of the radius. The velocity increases with decreasing density and
it is connected to the wind velocity at a prescribed point which we choose to be 10 km/s.  Once the velocity and 
density structures are specified, the radiation transport equation is solved consistently with the equations of 
statistical equilibrium and the energy balance as described in Hillier and Miller (1998). Several iterations of this 
procedure are performed until the density structure is consistent with the radiation force and the temperature 
distribution. The reference point at  $\tau_{Ross}$ = 10 is usually situated at V(r) $\sim$ 5 km/s, below  the 
connection point so the model parameters T$_{10}$ and $\log g$ apply to the underlying photosphere rather than the 
wind itself. 

The model of the star is specified by several parameters which can be combined into two groups. In the 
first group, there is the chemical composition, the atomic data and the stellar mass. For the analysis of
HD 5980, the values of these parameters are fixed for all epochs. The second group includes the stellar 
luminosity, the gravity acceleration, the temperature, the terminal velocity and the mass loss rate. 
For {\it Star C}, only one model fit was performed and we assume that all its derived parameters  remain 
constant over time.  For {\it Star A}, model fits were performed for each of the four spectra described in 
Section 2.

\subsection{{\it Star C} model}

\begin{figure}
\plotone{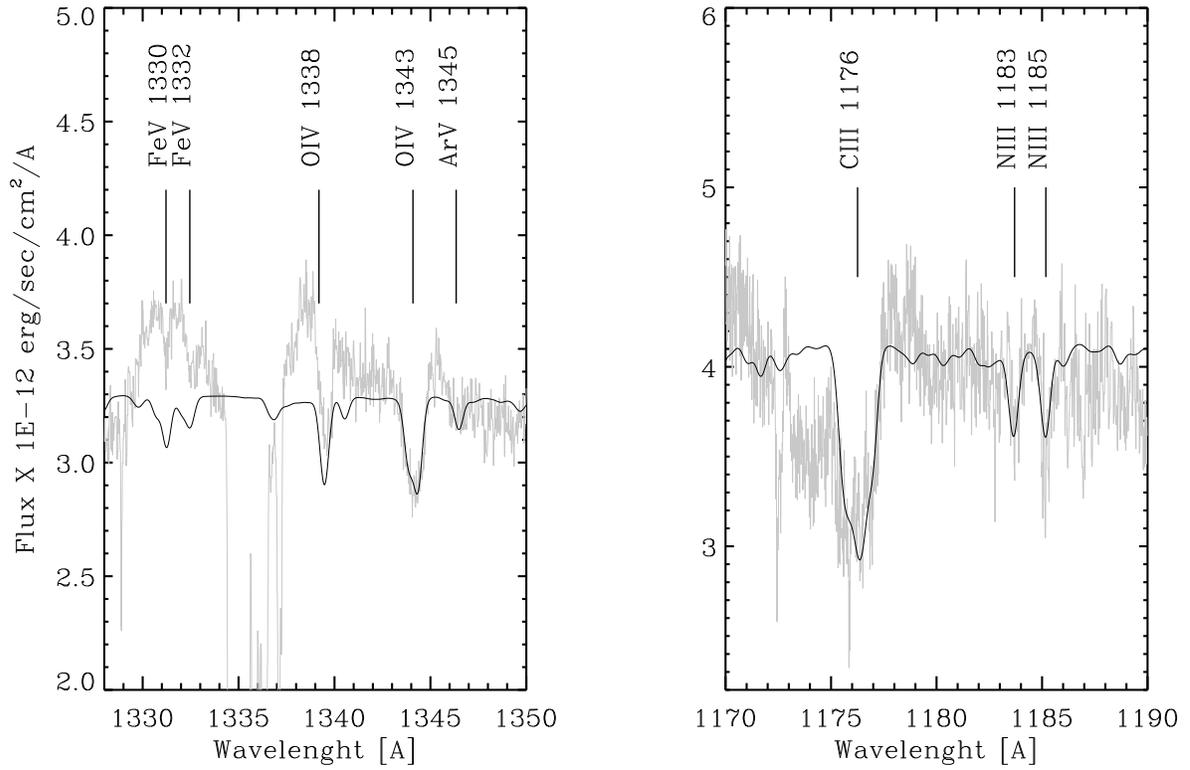}
\caption{Fragments of 2002 {\it HST/STIS} and {\it FUSE} spectra (grey line) compared with the
{\it Star C} model (black line). The model spectrum is shifted to the level of the observed
spectrum which is a sum of {\it Star A} and {\it Star C} fluxes.
\label{fig_starc}}
\end{figure}

\begin{figure}
\plotone{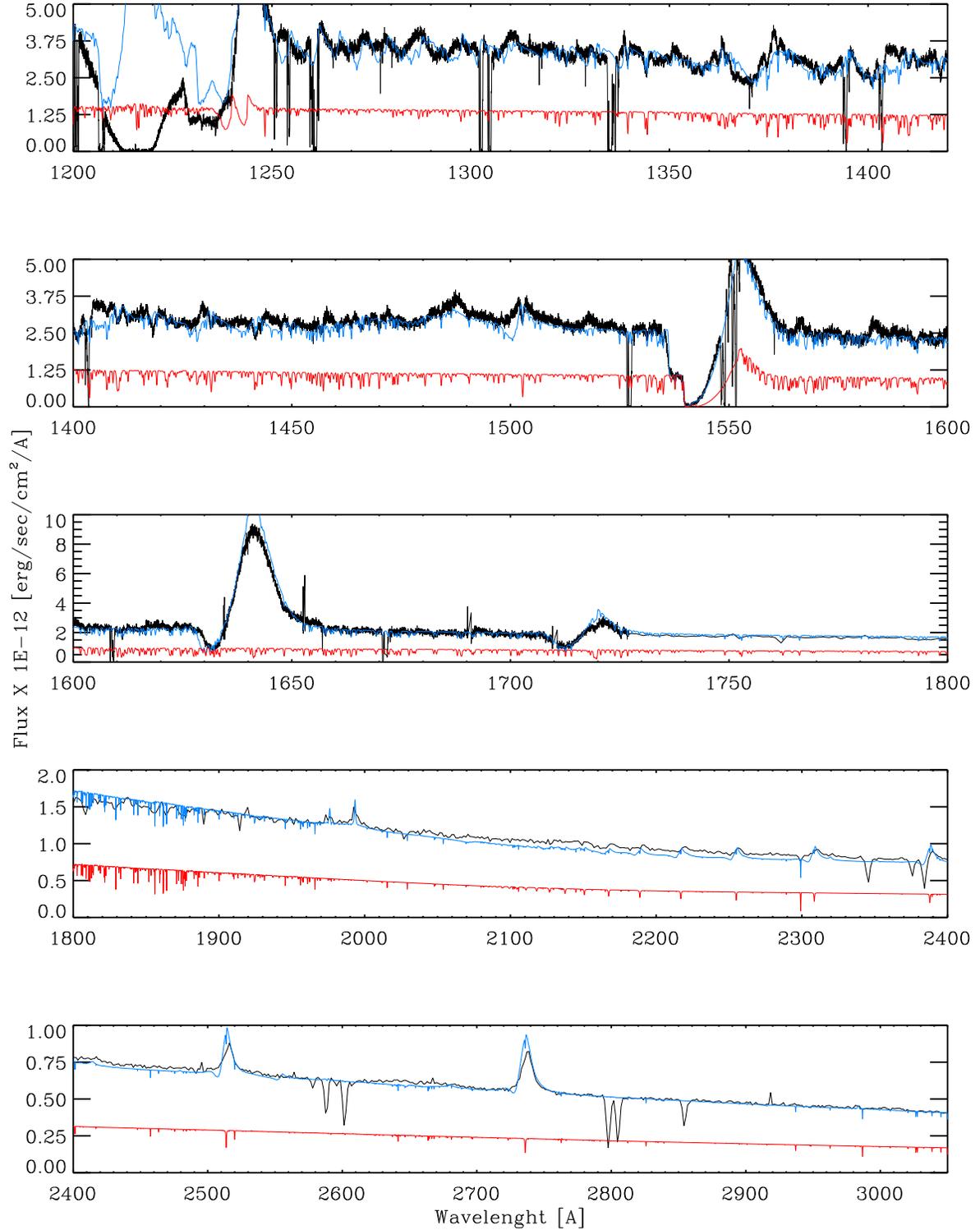}
\caption{Comparison between the {\it HST/STIS} spectrum obtained in September 2009 (black) and the sum 
of the current best model for {\it star A} at this epoch and {\it star C} model. The red line represents 
the computed spectrum of {\it star C} alone. Notice the coincidence of the {\it star C}'s continuum and 
the step in the blue wind of the \ion{C}{4} 1548/50 \AA\ P Cyg absorption. \label{fig_2009uv} }
\end{figure}

 Numerous photospheric absorption lines
belonging to {\it Star C} (Koenigsberger et al, 2002) are clearly present in the 2009 spectrum.
They were used to construct a model for this star.  Hunter et al (2007) found that Si and Mg composition in 
B-stars in NGC 346 is $\sim$ 0.2 of the solar abundance (Table~\ref{tab_params}). Assuming this is  representative for
all metals heavier than O, we adopted that chemical composition. The temperature of the model
was fixed by the \ion{O}{4} 1339-43 \AA\ and  \ion{O}{3} 5592 \AA\ lines. The model parameters were adjusted to attain
a reasonable fit to the depth of the  \ion{He}{1} and \ion{H}{1} optical absorption lines, clearly visible
on top of the emission lines. The wind velocity was set to the observed value of   $V_{black} = $ 1770 km/s
and the mass loss rate was restricted to the minimal value for which the \ion{C}{4} 1548-50 doublet is saturated.
The luminosity of {\it Star C} was fixed so that its continuum coincide with the ``step" of the observed
\ion{C}{4} 1548/50 doublet. We adopt  the CNO abundances of Sk 80 (Crowther et al. 2002), with which the lines of  
\ion{O}{3} 5592 \AA, \ion{C}{3} 1175 \AA\ and \ion{C}{3} 2297 \AA\ and \ion{N}{3} 1183-85 \AA\ (Fig.~\ref{fig_starc}) 
are adequately reproduced.  The adopted parameters are shown in Column 6 of  Table~\ref{tab_params} and the spectrum 
is shown in Fig.~\ref{fig_2009uv}. The model derived for  {\it Star C} does not
have significant emission lines in the UV-optical region other than \ion{C}{4} 1548-50 \AA, \ion{N}{4} 1718 \AA\
and \ion{N}{5} 1239-43 \AA. The comparison between the observed absorption lines and the computed
spectrum suggests a rotational velocity $V_{rot} \sin{i} = $ 80 $\pm$ 15 km/s and a radial velocity
$V_r \sim$ 60 km/s with respect to the {\it star A} + {\it star B} center of mass.   

It is important to note that Schweickhardt (2000) found periodic radial velocity variations in the photospheric
absorption associated with  {\it Star C}, with P$_{starC} \sim$96.5 days, and suggested that
it is itself a binary system.   Foellmi et al. (2008) also found RV variations in the   OIII 5592.4
line consistent with  Schweickhardt's (2000) conclusion. It is thus  important to keep in mind that our model 
for {\it Star C} corresponds to the combined spectra of two objects.  A second point to keep in mind is that
it remains to be determined whether this binary is gravitationally bound to the {\it Star A} $+$ {\it Star B} pair 
or whether it is merely a line-of-sight projection.  

\subsection{Star A}

In this section we describe the procedure for computing the model spectrum for {\it Star A}.  Because {\it Star C}
is in view at all times, the computed spectrum of {\it Star C} described above was added to all models of {\it Star A} 
before comparison with the observed spectrum. In the cases when only the normalized observed spectrum is available we  
compared the observations with a model flux calculated as

\begin{equation}
F\left(model\right)  = \frac{ \left(F_\lambda \left(starA\right) + F_\lambda \left(starC\right) \right)}{ \left(F_{cont} \left(starA\right) + F_{cont} \left(starC\right) \right)}
\end{equation}

\subsubsection{Chemical composition}

We constructed a fairly complex model including most of the important atoms in several ionization stages 
(Table~\ref{table_atoms}).  The main properties of the wind were obtained for each epoch.  If 
the observed spectrum at one epoch required an adjustment of the chemical composition, the change was made in all 
four epochs and the models were recalculated and the consistency checked. 

The He composition was determined by the decrement of the \ion{H}{1} and \ion{H}{1}+\ion{He}{2} optical lines. We 
fitted the lines in the spectrum obtained in 1994 and found that N(He)/N(H) = 1.0 $\pm$ 0.2 by number. This value
was fixed for all 4 spectra. Due to the significant abundance of  He, the mean atomic mass of the gas is larger than 
the Solar. In order to be able  compare {\it Star A}  with other objects having different He/H ratios we present 
the composition of all other elements by their mass fractions.

The carbon abundance was constrained mainly by \ion{C}{4} 5804/12\AA\ and \ion{C}{3} 1178 \AA\ as observed in 
the 2002 FUSE spectrum. The strong \ion{C}{4} 1548-50 doublet is not sensitive to the abundance. 
The \ion{N}{4} 7123 \AA\ line was used as a nitrogen  abundance indicator, and consistency was checked using
the UV N III, N IV and N V lines. Several iterations on all models 
at all epochs were made  until a consistent abundance of nitrogen was obtained. 
 
We do not observe any strong oxygen lines in the spectrum. The  \ion{O}{4} 1338-41 \AA\ observed in 
{\it STIS} spectra are formed in {\it Star C's} wind. The weakness of these lines in {\it Star A's} spectrum sets 
an upper limit to the O composition to 0.1  of the SMC value (1/50 of the solar composition). 
The phosphorus abundance was  set to 0.05 solar. The predicted \ion{P}{5} 1118-28 \AA\ lines based on this value 
agree with the 2002 observations.  \ion{S}{5} 1502 \AA\ line was fit with a sulfur abundance equal to  0.05 solar. 
The same abundance fits reasonably well \ion{S}{6} 933 \AA\ as well.  Aluminum abundance of 0.1 solar  fits \ion{Al}{3} 1855/63 \AA\ doublet observed in the 1994 spectrum.

Finally, we adopted Fe/H = 0.1 (Fe/H)$_\odot$ by mass since this abundance  reproduces well 
the \ion{Fe}{5} and \ion{Fe}{4} lines in the spectra of the 2000, 2002 and 2009 epochs. 
This iron abundance is lower than the one adopted for star C (Fe/H = 0.2 (Fe/H)$_\odot$ ) but we did not 
perform a rigorous analysis of the abundances of {\it star C} and we cannot exclude that it has the same Fe/H 
abundance $\sim$ 0.1(Fe/H)$_\odot$  as of {\it star A}. Nevertheless we kept the Fe/H abundance of {\it star C} 
to the commonly accepted value. The lower iron abundance of {\it star A} 
is consistent with the result derived from a comparison of the wind-eclipse effects in HD 5980 and the
Galactic WR system HD 90657 (Koenigsberger et al. 1987).\footnote{This result refers to the Fe abundance in
{\it Star B}'s wind, derived from a comparison between the Fe V pseudocontinuum and the N IV 1718 \AA\ line.}
For the elements Ne, Ar,  Cl, Ca, Cr, Mg and Ni, 
we adopted a 0.1 times Solar composition similar to the iron abundance.  The  spectrum does not show any observable spectral features arising from 
these species but they are important for the line blanketing and the radiation force and were included in the model.  

The composition of all other elements included in the model was also set to 0.10 of the solar value. The final 
chemical composition adopted for the four epochs is shown in Table~\ref{table_smc_comp}.  Note that the overall
abundance for {\it Star A} is 0.1 Solar while that which was used for the {\it Star C} model is 0.2 Solar.  The
uncertainties in the model fits lead to uncertaintites in the chemical compositions of $\pm$0.1 dex, so that these
two metalicity values are consistent, within the uncertainties. 
\subsubsection{Wind velocity}

\begin{figure}
\plotone{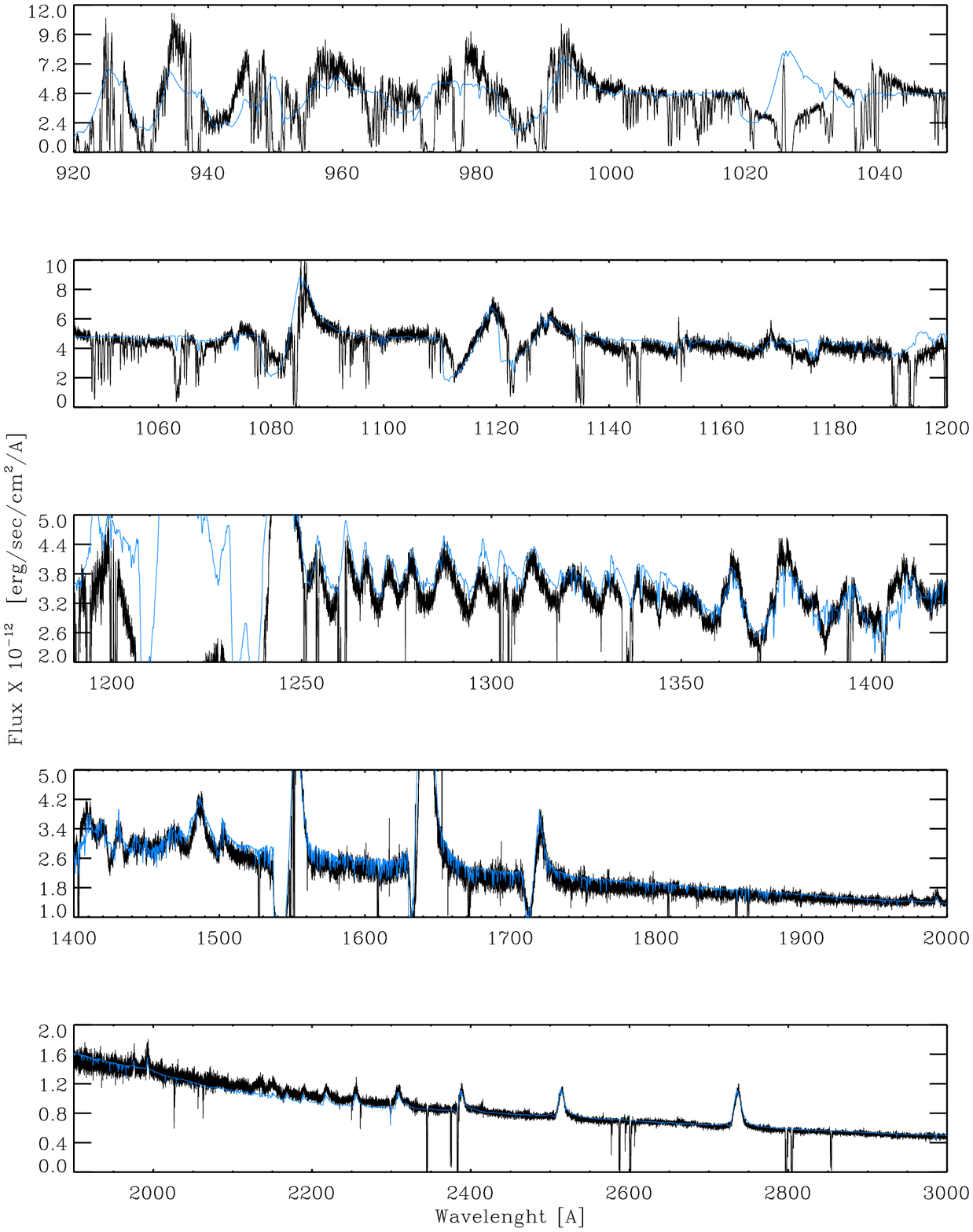}
\caption{As if Fig.~\ref{fig_1994} but for the  April 2002 epoch.
\label{fig_2002} }
\end{figure}

The calculated spectrum of a model is sensitive to several parameters. The first two parameters
are the wind velocity law and the terminal speed, $V_\infty$.  The Fe V and Fe VI lines in
the $\lambda\lambda$1270--1450 \AA\ region are optically thinner than the stronger lines present
in the spectrum, and thus they are more sensitive to properties of the inner wind regions, particularly
the velocity law.  We find that a $\beta\geq$2 velocity law adequately reproduces their line-profiles.
Values of $\beta<$2 produce profiles with a more ``box-like" shape than observed. Thus, we set
$\beta=$2. The value of V$_\infty$  was initially chosen for each epoch from Table~\ref{tab_star_velocities}. However, we
found  it challenging to  achieve a good fit to the P~Cygni absorption components of all lines with 
the same value of V$_\infty$.  In general, the value of V$_\infty$ deduced from  the C IV line leads
to a model in which the extent of the \ion{He}{2} 1640 \AA\ and \ion{P}{5} 1118-28 \AA\ P~Cygni 
absorption is greater than observed  (Fig.~\ref{fig_2002}). The difference is $\sim$ 500 km/s. As
mentioned in Section 2, the wind velocity determined from the He II 1640 P~Cygni edge is systematically
smaller than that derived from  \ion{C}{4} (listed in Table~\ref{tab_star_velocities}).   A slower wind velocity law (i.e. $\beta>$2) 
reduces this discrepancy.  However, in the case of \ion{N}{5} 1240 \AA\, our models require a faster $V_\infty$
than that derived from \ion{C}{4}, although the measurements of Table~\ref{tab_uvspec2} give values that are consistent,
within the uncertainties, with \ion{C}{4}.  Increasing the value of $\beta$ does not eliminate this inconsistency
in the model. Similar discrepancies are  observed in some  central stars of planetary nebulae (Morisset and Georgiev 2009; 
Herald et al. 2011; Arrieta et al. 2011). This phenomenon needs further investigation. For the purposes of 
this paper we used the V$_\infty$ as derived from the \ion{C}{4} 1548-50 \AA\ line and $\beta$ = 2.
 
\subsubsection{Stellar mass and luminosity}

No  feature in the spectrum is clearly dependent on the gravity acceleration. This is not surprising given  
that the continumm forming region extends beyond the photosphere. But the mass of the star, M\,$\sin i$, $i$ being the 
orbital inclination, has been estimated from the analysis of the light curve and radial velocity curve to be 
between 60 and 80 M$_\odot$. Our modeling shows that at the observed luminosities and mass loss rates, a star with a
mass smaller  than 90 M$_\odot$ is above the Eddington  limit. To keep the star stable, we assume a mass of 90M$_\odot$ 
which, within the errors,  is consistent with the orbital solution and keeps the stellar photosphere below 
 the Eddington limit. The $\log g$ was adjusted so the mass of the star is maintained the same for all epochs. 
 
The luminosity at different epochs was determined from the continuum level of the flux calibrated UV spectra. We assume
a distance to the SMC of 64 kpc which is an average of the values obtained by Hildich et al. (2005) and North et al. (2010).
The reddening was determined by  comparing  the model and observed spectral energy distribution in the  
920 \AA\ - 11500 \AA\ range and using the Cardelli et al. (1989) reddening law. We obtained a good fit with  
E(B-V)=0.065$\pm$0.005 and R = A$_V$/E(B-V) = 3.1 which was used for all epochs.

\subsubsection{Temperature, mass-loss rate and clumping}

The temperature and the mass loss rate are the most difficult parameters to determine. The change in the spectrum 
over the different epochs was so great  that one cannot use the same diagnostic feature for the analysis for all
epochs. The ever present helium lines are affected by  {\it Star B} in a yet unclear way so we avoided using them. 
We concentrated on  lines which  significantly change from epoch to epoch and therefore can safely be assumed to 
arise in {\it Star A}. Nitrogen is present as \ion{N}{3}, \ion{N}{4} and \ion{N}{5} in different epochs and due 
to this variability we deduce that Nitrogen lines are formed mostly in {\it Star A}'s wind.  We used the 
\ion{N}{4} 1486] line as a mass loss diagnostic and the \ion{N}{3} 1750 \AA\ and 4640 \AA\ lines (when present) 
were used to constrain the  temperature. Although we did not use the helium lines for this analysis,  the fits to 
the \ion{He}{2} lines are reasonably good in most of the epochs. \ion{He}{1} lines are reasonably fit in 1994 and 2000 epochs
but the models underestimate them in 2009. The profile of \ion{He}{1} lines is also narrower for the observed V$_\infty$.
This suggests that at least part of the observed \ion{He}{1} 5876 \AA\ emission originates in wind-wind interaction region.    

It is now well established that the winds of the massive stars are not homogeneous but rather made of clumps, although the 
true nature of the clumping is not known. The main effect of the clumping in the WR winds is the absence of  electron 
scattering wings of the strong emission lines. Following the same procedure as in Hillier et al. (2003) we used the 
volume filling factor $f$ prescription. We adjusted $f$ until we fit the red wing of \ion{He}{2} 4686 \AA \, line and 
then check the consistency with the other strong \ion{H}{1} + \ion{He}{2} lines. The observed electron scattering 
wings are very weak which restrict the volume filling factor to $f \le$  0.025. This is an upper limit to $f$. 
In some spectra the wings of the lines are even weaker,  but smaller values for $f$ might be beyond the validity 
of the approximations used in the formalism.  We used $f$=0.025 for all models for all epochs.  

\subsubsection{The individual spectral fits}

\noindent {\bf Spectrum 1994:}
Of the four spectra, this one corresponds to the coolest temperature.  \ion{N}{3} 1750\AA\ blend is clearly seen and we 
use the ratio \ion{N}{3} 1750/\ion{N}{4} 1718\AA\  as a temperature diagnostic. The optical part of the spectrum shows 
\ion{N}{3} 4640\,\AA\ line which is also well reproduced. In the low resolution {\it IUE} spectrum, the observed 
\ion{Fe}{4} lines in the 1600-1700\AA\  are much weaker that  predicted. 
The mass loss rate was fixed mainly by \ion{N}{4}] 1486 but  is consistent with the other UV emission lines.  
Taking into account the clumping, our derived $\dot{M}$ is in reasonable agreement with that derived in Koenigsberger et al. (1998b) for the same 
observed spectrum.  A likely explanation for this discrepancy is that the model used in that study considered
only hydrogen and helium, thus missing the important influence of the line blanketing. 

\noindent{\bf Spectrum 2000:} The temperature of the star is low enough to show a measurable \ion{N}{3} 4640\AA. 
We used that line to restrict the upper limit of the temperature.  The mass loss rate was constrained by 
\ion{N}{4}] 1486 \AA\ and further check with the \ion{Fe}{5} UV complex. The model also reproduces well the 
optical \ion{He}{1} and \ion{He}{2} lines. 

\noindent{\bf Spectrum 2002:} The spectrum of this epoch does not include the optical wavelengths. We set the 
temperature based on the \ion{Fe}{6}/\ion{Fe}{5} features observed between 1250\,\AA\ and 1400\,\AA\ and 
1420\,\AA\ and 1480\,\AA\ respectively. The mass loss rate is derived from  
\ion{N}{4}] 1486 and  the \ion{P}{5} 1118-28 \AA\ doublet. 

\noindent {\bf Spectrum 2009:} The temperature was limited from below by the absence of \ion{N}{3} 4640 \AA. 
This line is present in the previous epochs but is weak in  this last spectrum. In addition  we used the 
\ion{Fe}{5}/\ion{Fe}{4} UV complex.  

The uncertainties in the results obtained from the model fits are difficult to estimate
due to the large number of free parameters which are not completely independent. However, the models show that
changing T($\tau_{Ross}$=10)  by more than $\sim$2000K causes   changes in the computed spectrum that make it noticeably
different from the observations.  In an analogous manner, the uncertainty in the mass-loss rate can be   
estimated to be 30\%.  The uncertainty in the luminosity due to the uncertainties in the flux calibration and 
in the S/N of the spectra is $\lessapprox$20\%. The absolute uncertainty in the parameters is much larger due 
to unclear correlations between them which  are difficult to quantify. We  minimize these errors by making our models
for the different epochs consistent so the differences in the parameter values are expected to reflect  real
changes in the stellar physics.

\subsection{Results}

\begin{figure}
\plotone{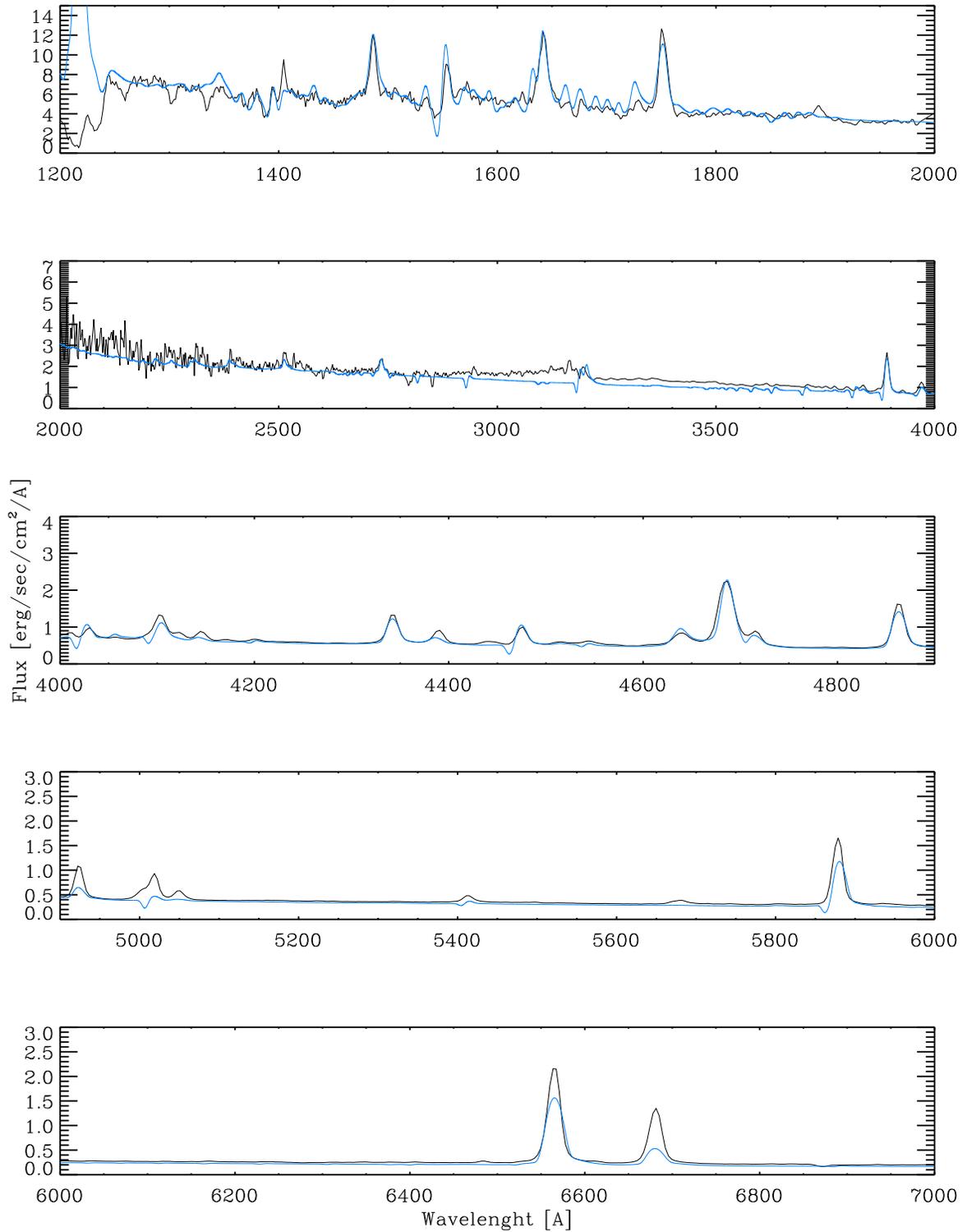}
\caption{Comparison between the spectrum obtained on December  30,  1994 (black)  and the sum of the 
current best model for {\it star A} at this epoch and {\it star C} model (blue). The singlet \ion{He}{1} 
lines are underestimated while the triplet lines are well reproduced.
\label{fig_1994} }
\end{figure}

\begin{figure}
\plotone{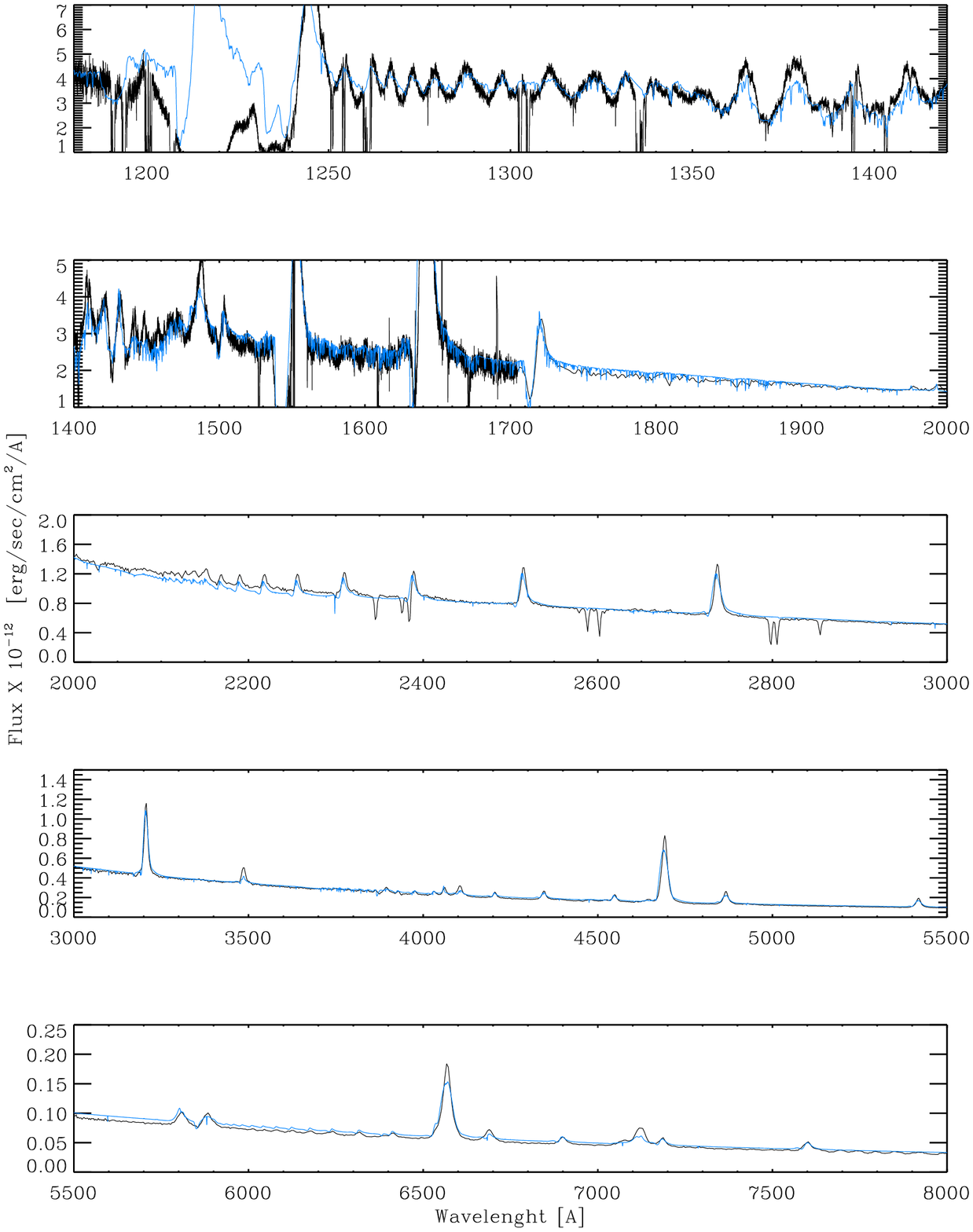}
\caption{As if Fig.~\ref{fig_1994} but for the  April 2000 epoch.
\label{fig_2000}}
\end{figure}

\begin{figure}
\plotone{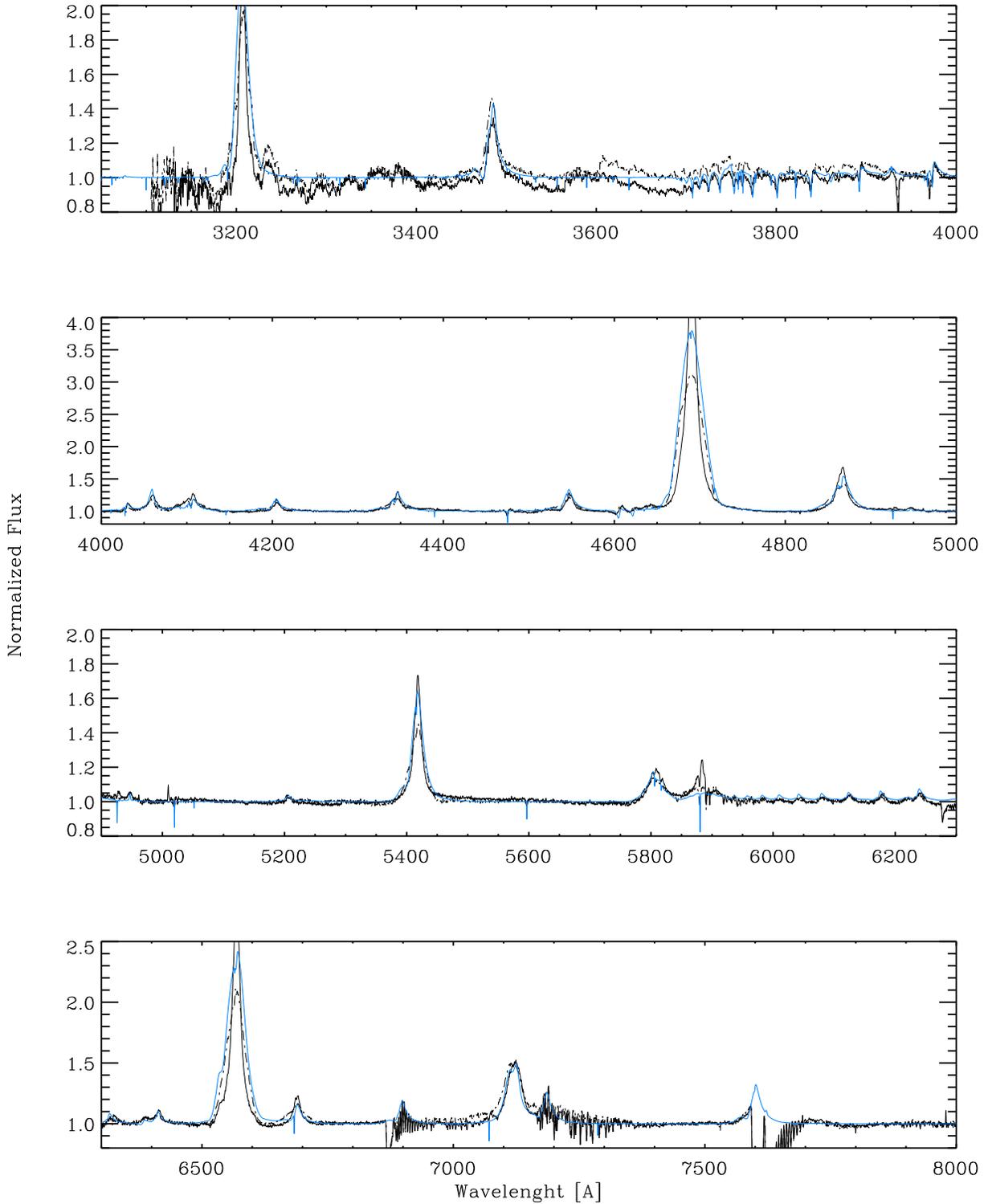}
\caption{As in Fig.~\ref{fig_2009uv} but for the optical region. The spectrum is affected by the presence 
of strong telluric absorption lines in 6850\AA, 7200\AA\ and 7550\AA.  The observed profile of \ion{He}{2} 
4686\AA\ and H$_\alpha$ is much narrower than the model spectrum. Breysacher et al. (1982) showed that 
\ion{He}{2} 4686 \AA\  changes its profile with the orbital period being narrower at the eclipses. The model 
show that the line width outside the eclipse correspond to the UV P Cyg lines and there is missing emission 
during the eclipse. The interpretation of this effect requires a reliable model for {\it star B} which will be 
treated in a forthcoming  paper. 
\label{fig_2009opt} }
\end{figure}

Table~\ref{tab_params} summarizes the parameters derived from the models of {\it Star A} in the epochs 1994, 1999, 2002 and 2009 as
well as the model of {\it Star C}.  The comparison between the observations and the calculated spectra are shown
in Figs.~\ref{fig_2009uv}, ~\ref{fig_2002}, ~\ref{fig_1994}, ~\ref{fig_2000} and ~\ref{fig_2009opt}.

Three different radii are listed in Table~\ref{tab_params}: 1) R$_{10}$, which is the radius at which the Rosseland optical
depth $\tau_{Ross}$ = 10, and it is this reference radius at which the gravity acceleration is $\log{g}$
and where the temperature T$_{10}$, is given by  the CMFGEN model; 2) R$_s$, which  is the radius
at the sonic point, were the optical depth and temperature are, respectively, $\tau_s$ and  T$_s$;
we define the photosphere as that  region of the star with r $<$ R$_s$ and the wind where r $>$ R$_s$;
3) R$_{2/3}$, which is the radius  of the continuum-forming region, where $\tau_{Ross}$ = 2/3, and
here T$_{eff}$ = T(R$_{2/3})$.

\begin{figure}
\plotone{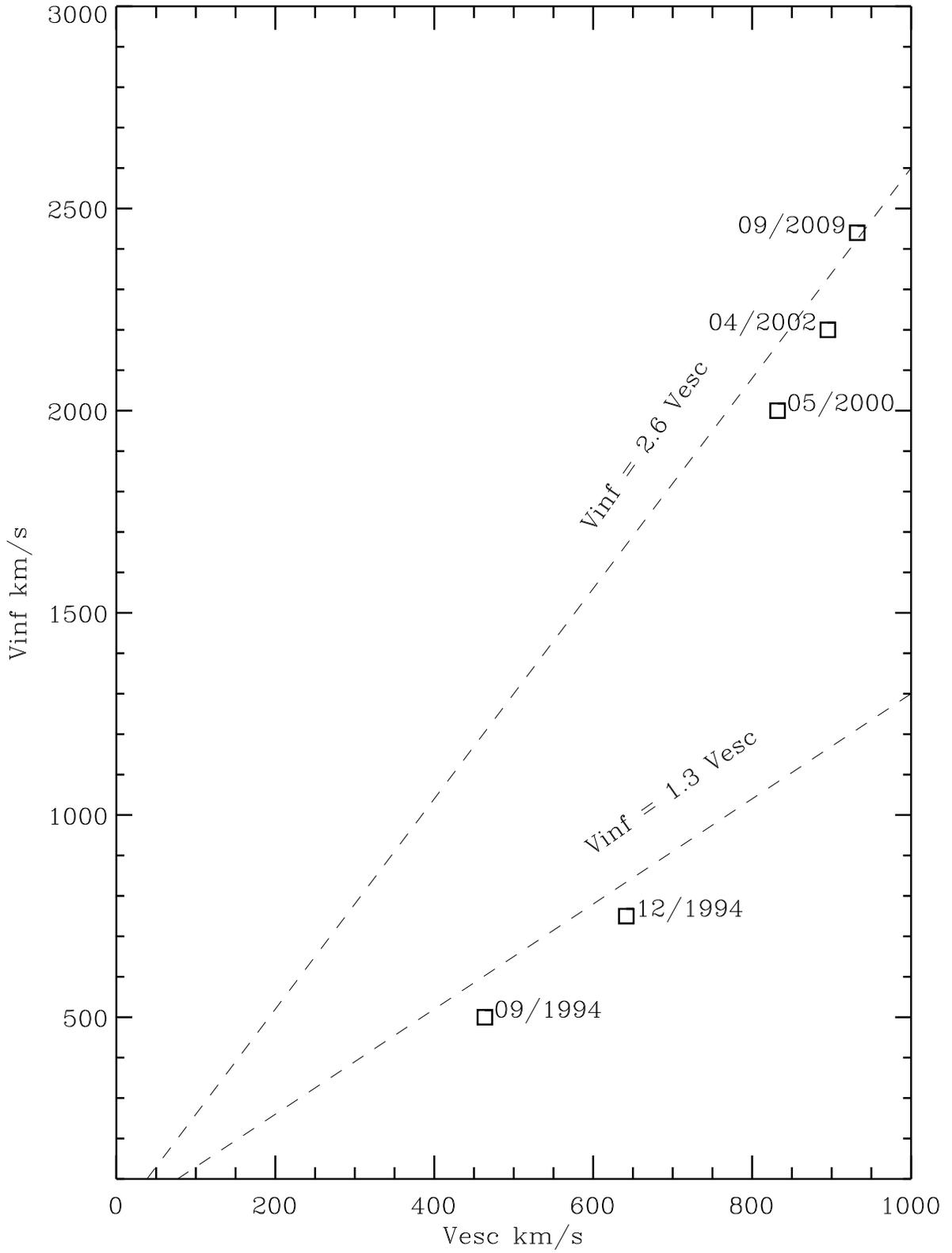}
\caption{The relation between V$_\infty$ and V$_{esc}$ at different epochs.
\label{fig_vesc}}
\end{figure}

The most outstanding results derived from these models are:

\begin{enumerate}
\item{} The luminosity of {\it Star A} is higher during the 1994 eruptive event than at later epochs. This is 
consistent with the results obtained by Drissen et al. (2001) from a spectrum take on 1994 September 21. Their
luminosity is even higher than our value for the spectrum obtained on 1994 December 31.  The luminosity is lowest 
in 2000 April,  and then rises slightly  in 2002 remaining at the same level until 2009.  Thus,   we are
now able to confirm that the 1994 eruption is an event  which did not occur at constant  luminosity.
 
\item{} {\it Star A}'s mass-loss rate is a factor of $\sim$4 larger in 1994 than at later epochs. 
The larger luminosity most likely plays an important role in driving the larger mass-loss rate.

\item{} The radius at $\tau_{Ross}$=2/3, $R_{2/3}$=124 $R_\odot$, is larger in 1994 than at any other epoch, and
it is very much larger than $R_s$=21.5 $R_\odot$, the radius at the sonic point in 1994. The continuum-forming
region  is very extended. Fig.~\ref {fig_temp} shows the temperature structure of the 1994 model
where one can also see that due to the larger wind density, the temperature is significantly higher
throughout the wind.  It is also quite remarkable that the temperature at the sonic point is $T$=10$^5$ K. This 
high temperature is  due to the fact that radiation is trapped within the extended continuum emitting region.

\item{} $R_{2/3}$ declines  after 1994  reaching 28 R$_\odot$ in 2009.  This  value is   
larger than the size of the observed continuum-emitting region of {\it Star A} in 
1979 (22--25 R$_\odot$; Perrier et al. 2009), consistent with the notion that {\it Star A} has still 
not recovered its minimum state of activity.  The decrease in $R_{2/3}$ reflects the shrinking size of  
the continuum forming  region as the density drops. Note, however, that $R_s$=19.5 $R_\odot $ in 2009 epoch, indicating
that the continuum formation region is still rather extended.

\item{} The  most significant change in R$_s$ occurred in the 2000 epoch, at which time
it was larger than  during the other epochs. This is consistent with the long-term behavior of the optical
light curve which peaked at around the 2000 epoch  and  with the notion that the long-term activity cycle
and the brief eruptions of 1993-1994 seem to be separate phenomena (Koenigsberger et al. 2010).

\item{} There is a  30\% variation  in $\dot{M}$ between 2000 and 2002 and  it remains relatively
constant thereafter.  Although it is tempting to suggest a declining trend, the variation
lies within the uncertainties of the model results.  Hence, we tentatively conclude  that $\dot{M}$
is approximately constant over the timescale 2000--2009.   This leads to the conclusion
that the observed spectral changes  during this timescale result primarily from the  declining  wind density,
caused by the  increasing wind velocity.  Hence, we are now in the possibility of understanding the relation 
between the flux in \ion{N}{4}] 1486 \AA\ and the terminal wind speed (Fig.~\ref{fig_niv_vel}). 
Based on the models we can conclude that the reduction of the post-eruption line intensities is caused mainly
by the increase of V$_\infty$.   Assuming that V$_\infty$/V$_{esc}$ is constant (Fig.~\ref{fig_vesc}) one can
conclude that the changes in V$_\infty$  are caused by  a decreasing radius of the star (Table~\ref{tab_params}).  The 
relation between  \ion{N}{4}] 1486 \AA\ and the terminal velocity is maintained during
the pre-eruption epochs and it follows the same general trend (Koenigsberger et al., 2010) but in the opposite
direction.  V$_\infty$ decreases while \ion{N}{4}] 1486 \AA\ increases which points to an increase in the stellar radius.
This leads us to conclude that the S Dor-type $\sim$40 year variability timescale of {\it Star A} is due 
to variations in its radius.

\end{enumerate}

The S Dor type variability of {\it Star A} is similar to that of AG Car (Groh et al. 2009) and S Dor (Lamers 1995).
Excluding the 1994 eruption event, the maximum of the optical brightness in HD 5980 was attained between 1996-2000. 
At that time, the terminal wind velocity and the temperature were lower  than in 2009, when the visual brightness 
was approaching its  minimum.  In this sense, during its long term variability, {\it Star A}  behaves similarly to 
AG Car and S Dor. Lamers (1995) explained this variability as a pulsation of  10$^{-3}$ to 10$^{-2}$ of the stellar 
mass in S Dor, and Groh et al. (2009) reached the same conclusion for AG Car.  Lamers (1995) suggested that the larger 
radius decreases the gravity acceleration and increases the  scale height which leads to larger mass loss.  He also 
suggested that the luminosity of the star should be lower during the maximum of the optical brightness due to the work 
the star must perform to lift the upper layers.  In agreement with this prediction we detected a small difference in 
the luminosity in 2000 (close to maximum) and 2009 (close to minimum) epochs which add weight to the suggestion that 
S Dor variability of {\it star A} is caused by a pulsation.  The exact physical mechanism causing the pulsation is  
unclear but now that we have three objects showing similar effects,  the question arises as to whether this may be 
a  common phenomenon in this class of objects.

\section{The 1994 eruption and the {\bf bi-stability jump of a second kind}}

\begin{figure}
\plotone{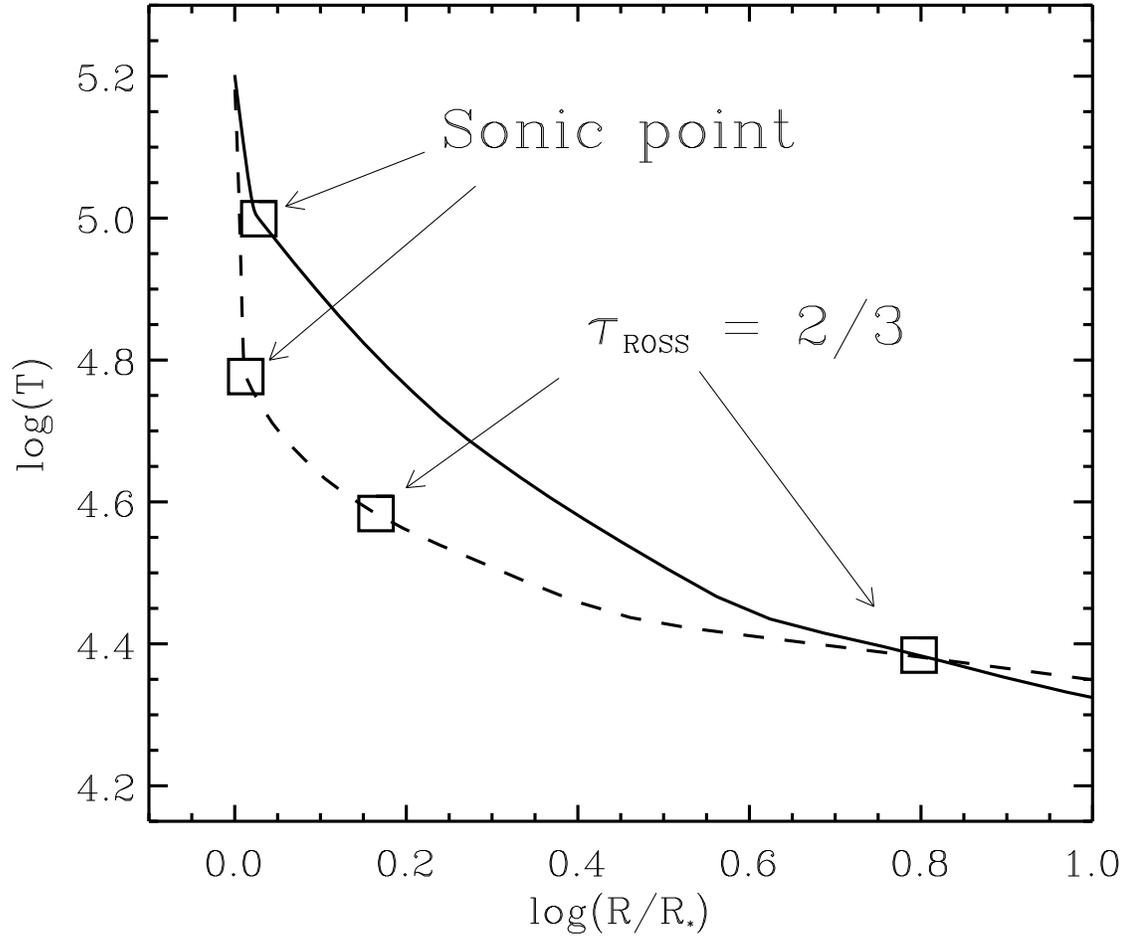}
\caption{Temperature distribution in the 1994 (solid line) and 2009 (dashed line) models. The positions of
the sonic point and the continuum forming regions ($\tau_{Ross}$ = 2/3) are indicated.  The lower mass
loss rate in the 2009 model moves the  continuum-forming region inward.
\label{fig_temp}}
\end{figure}

The results presented in Table~\ref{tab_params} indicate that the wind structure of {\it Star A} in 1994
was significantly different from its  wind structure during the other 3  epochs.
This difference is best illustrated in Fig.~\ref{fig_temp} where we plot  the  temperature structure derived
from the spectra of 1994 and 2009. The temperature is higher throughout most of the wind in 1994 than in 2009,
even though the observed  1994 spectrum looks cooler.
This is a consequence of the higher density in 1994 which makes the  wind more optically thick.  

The velocity structures are also different, in particular, the relation between the 
terminal wind velocity, V$_\infty$ and the escape velocity V$_{esc}$, which is  
defined as

\begin{equation}
V_{esc} = \left(\frac{GM\left(1-\Gamma\right)}{R_{10}}\right)^{1/2}
\end{equation}

\noindent where M is the stellar mass assumed to be 90 M$_\odot$. The ratio between gravity acceleration and the
radiation pressure $\Gamma = g_{el}/g_{grav}$ takes into account only the radiation pressure caused by electron
scattering (Lamers and Cassineli, 1999). We calculate V$_{esc}$ using $\Gamma$ as determined by the CMFGEN model
at R$_{10}$.   Fig.~\ref{fig_vesc} shows  V$_\infty$ plotted against  V$_{esc}$.  The values for the three
post-eruption epochs (2000, 2002 and 2009) lie close to a single straight line, the slope of which is $\sim$ 2.3.
The wind during the 1994 epoch is different.  In this case,  V$_\infty$/V$_{esc} ~\sim$ 1.2.  The velocities obtained
by the Drissen at al. (2001)  for 1994 September lie along the same relation. 

The change in the V$_\infty$/V$_{esc}$ relation together with the cooler effective temperature of the 1994 spectrum
are  strongly suggestive of a phenomenon associated with  the bi-stability jump (Lamers et al. 1995). The bi-stability jump
occurs  when the temperature in the inner wind decreases to a value at which the ionization fraction of Fe$^{2+}$ increases
leading to changes in the line opacities and a changes of the driving force, higher mass loss rate and reduce in 
V$_\infty$/V$_{esc}$ from $\sim$2.6 to 1.3 (Vink et al, 1999). This occurs at $T_{eff}\sim$21000 K. 

This is remarkably similar to what we find for HD 5980.  There is, however,
a major difference between the cause of the bi-stability jump  observed in O and B stars and the
phenomenon that our models indicate is occurring in  HD 5980.  As was shown by Lamers and Cassinelli, (1999) 
and Vink et al. (1999), the mass loss rate is defined by the force below the sonic point. In the optically 
thin O and B star's winds the sonic point is above the photosphere, thus the condition in it can be described 
by the effective temperature. On the other hand, the sonic point of the optically thick wind of HD5980 is 
located below the pseudo photosphere (Fig.~\ref{fig_temp}). In the 1994 model, the point at which the 
temperature T$\sim$T$_{eff}\sim$21000 K lies at $\sim$6 $R_s$ far above the sonic point. 
Thus, changes in the ionization structure at a distance of $\sim$6 $R_s$ have little
effect on the overall wind structure.  

In search of an alternative cause of the jump we followed the reasoning of Lamers and Nugis (2002), and  
Gr\"{a}fener and  Hamman, (2008). These authors showed that the WR winds below the sonic point are driven 
not by the line force but rather by the radiation pressure due to continuum opacities. We applied that idea to 
{\it star A}'s wind. In Table~\ref{tab_params} we see that the optical depth of the sonic point $\tau_s >$ 1 
for all four epochs, indicating that the wind is driven primarily by the continuum force.  Lamers and 
Nugis (2002), and  Gr\"{a}fener and  Hamman, (2008) showed that in order to drive a WR-type wind, the continuum 
opacity (sum of the true continuum opacities and the pseudo continuum of overlapping lines) has to be dominated by one 
of the bumps in the iron opacity.  Gr\"{a}fener and Hamman, (2008) 
showed that the required high mass loss rate is obtained only if  the temperature at the sonic point is 
either in the 30-70kK range (``cool" bump)  or around 160kK (``hot" bump). Table~\ref{tab_params} shows that 
all of the post eruption models have T$_s \sim$~48--50 kK, well within the first of these temperature intervals.  
One can thus expect that at almost constant luminosity the  mass loss rate will be almost constant,
which is indeed the result from our models (Table~\ref{tab_params}).    During the eruption the luminosity of the star 
increased.  The higher luminosity heated the inner part of the wind changing the opacity contributors 
from those of the  ``cool" bump to those of the ``hot" bump, thus changing the driving force and therefore 
the mass loss rate.  Our model for 1994 December gives  T$_s$ = 100kK (Table~\ref{tab_params}).   This temperature 
is too high for the ``cool" iron opacity bump but too cool for the ``hot" bump at 160kK. Gr\"{a}fener and Hamman 2008) 
suggested that winds with T$_s$ in the intermediate temperature range are unstable. We speculate that at the onset 
of the eruption T$_s$ was high enough and the very dense wind was  driven by the high temperature iron bump.  As the 
luminosity decreased, the  cooler wind became unstable  and it rapidly switched back to the ``cool" iron bump driving 
with T$_s$ around 50 kK. 

We suggest that  a bi-stability jump of a second kind operates in {\it Star A}. The 
sudden change in the wind properties is due to a transition from the ``hot" to the ``cool" Fe-bump 
opacity regimes mechanism that was shown by Lamers \& Nugis (2002) and Gr\"afener \& Hamann (2008) 
to be needed to drive the winds of WR stars.

Our models show that the increased luminosity can  indeed change T$_s$, changing the entire wind structure during the eruption. 
 Fig.~\ref{fig_temp} shows the change in the temperature structure of the wind in the 1994 and 2009 models. In  the less dense wind of 
2009, the continuum forming region has moved further in and T$_{eff}$ increased. The sonic point is 
at a similar radius but due to the overall smaller temperature  gradient it is significantly cooler.
 In addition to the changes in the temperature,  the value of $\Gamma$   changes from $\sim$ 0.75 during the eruption to 
$\sim$ 0.53 afterward.  A similar result was found by Smith et al. (2004). These authors showed that when the star approaches the 
Eddington limit, the optical depth of the sonic point increases and it moves below the photosphere. The increased luminosity of {\it star A}
moved it closer to the Eddington limit and the wind reacted in a similar manner. Furthermore
Gr\"afener and Hamann (2008) found that the WN stars with SMC chemical composition require 
$\Gamma>$ 0.75 to maintain a high mass loss rate.  In addition  the Vink et al. (2011) models show a transition  
from O star wind to WR wind when $\Gamma$ increases above 0.7. Based on these considerations we can conclude 
that the wind during the eruption is governed by different forces then the wind  in the post eruption epochs.  
We speculate that during the eruption, the high luminosity heats the inner part of the wind, driving much higher 
mass loss trough the ``hot" iron opacity bump.  After the eruption the temperature at the sonic point drops 
making the wind unstable. 
The mass loss rate drops, which causes further cooling of the sonic point until it reaches the ``cool" iron bump where 
the wind stabilizes again and the star continues its long-term variability but with a restructured wind.  

Hence, we  suggest that  for the dense winds such as those in HD 5980, there exists a second type of bi-stability jump 
than that proposed by Lamers et al. (1995). It is caused by transition between the ``hot" and ``cold" iron opacity bump 
and is correlated mainly with the stellar luminosity rather than with the effective temperature.  Smith et al (2004) 
discussed the position of the sonic point in the typical LBVs. They showed that for the derived parameters of most of 
the LBVs, the sonic point is located above the photosphere and thus their winds change according to the ``classical " 
bi-stability jump. Here we speculate that the most luminous LBVs, such as {\it star A}, have optically thick winds with 
pseudo photospheres  above their sonic points even on the hotter side of the S Dor instability strip. Thus their 
excursion to the cool branch of the strip is coused by a second kind of bi-stability jump related to the changes in 
their continuum opacities.

\section{Conclusions}

HD 5980 is a relatively nearby system in which processes involving eruptive phenomena and massive star evolution in binary 
systems in a low metallicity environment can be studied.  In this respect, is important to gain an understanding of the
phenomena that it has presented over  the past few decades.

In this paper we present the results of the analysis of a  collection of UV  and optical spectra obtained 
between 1979 and 2009 and performed CMFGEN model fits to spectra of 1994, 2000, 2002 and 2009.  The primary results 
are as follows:
a) The long term S Dor-type variability is associated with changes of the hydrostatic radius.  We find that the radius
of the eruptor at the sonic point was largest in the year 2000 (R$_s$=24.3 R$_\odot$), coinciding with the long-term 
maximum in the visual light curve.  The minimum value in our data set is in 2009 (R$_s$=19.6 R$_\odot$).

b) The  1994 eruption involved changes in the eruptor's  bolometric luminosity and wind structure.  The luminosity
was larger by at least 50\% with respect to the subsequent epochs and the temperature throughout most of the wind
was significantly higher.  The latter is a consequence of the very large optical depth which causes photons to be
trapped within this wind region.  At the same time, the velocity structure was one in which the relationship between
the terminal wind velocity and the escape velocity  V$_\infty$/V$_{esc}\sim$1.2, contrary to other epochs in which
this ratio was $\sim$2.3.
c) the emission-line strength, the wind velocity and the continuum luminosity underwent  correlated
variations in the sense that a decreasing V$_\infty$ was associated with  increasing  emission line and
continuum levels. These correlations can be understood in the context of the increasing size of the hydrostatic stellar
radius.   
d) The spectrum of the third star in the system ({\it Star C}) is well-fit by a T$_{eff}$=32 K model atmosphere 
with SMC chemical abundances.  The abundances of the eruptor, {\it Star A}, show He and N enhancements, roughly
consistent with the CNO cycle equilibrium values.  A similar composition was  obtained by Groh et al. (2009)
for AG Car and Hillier et al. (2001) for $\eta$ Car.

For all epochs, the  wind of the erupting star is optically thick at the sonic point and is thus driven mainly 
by the continuum opacity.  We speculate that the wind switches between two stable regimes
driven by the ``hot" (during the eruption) and the ``cool" (post-eruption) iron opacity bumps as defined
by Lamers \& Nugis (2002) and Gr\"afener and Hamann (2008), and thus the wind may undergo a bi-stability jump
of a different nature  from that which occurs in OB-stars.

\section{Acknowledgements}

We thank Alfredo D\'{\i}az and Ulises Amaya for computing assistance.
This research was supported under UNAM/DGAPA/PAPIIT grants
IN106798, IN123309  and CONACYT grants 48929 and 83016. 
LNG acknowledges the support from CONACyT  grant 141530. RHB acknowledges partial support from ULS DIULS grant, and DJH acknowledges support from grant  HST-GO-11623.01-A and HST-GO-11756.01.
The HST is operated by the STSCI, under contract with AURA.

{}
\clearpage


\begin{deluxetable}{lllllllll}

\tablecaption{High dispersion spectra \label{tab_uvspec1}}     
\tablecolumns{9}                                       
\tablewidth{0pt}
\tablehead{
\colhead{Spectrum} & \colhead{HJD\tablenotemark{1}} & \colhead{Phase} & \colhead{FES mag} & \colhead{HeII$_{edge}\tablenotemark{2}$}
&\colhead{NV$_{sat}$\tablenotemark{3}}& \colhead{F(N IV])\tablenotemark{4}}&\colhead{F(NIII)\tablenotemark{4}}& \colhead{F(1850\AA)\tablenotemark{5}} 
}
\startdata
 4277     &43921.1  &0.57   &11.68   &-2770         &-2800      &$<$ 1.  & 1.0   &1.47      \\   
 4345     &43928.6  &0.91   &11.69   &-2620         &-2760      &$<$ 1.  &$<$1   &1.53      \\       
 4958     &43981.0  &0.68   &11.72   &-2800         &-2050      & $<$ 1. &$<$1   &1.43      \\      
          & \nodata &\nodata&\nodata &\nodata       &-2790      &\nodata &\nodata&\nodata   \\
11175     &44632.3  &0.48   &11.61   &-2920         &-2610      &2.8     &2.7    &1.58      \\   
11190     &44634.6  &0.60   &11.63   &-2920         &-2540      &$<$1.3 &$<$1   &1.64      \\
15072     &44869.5  &0.80   &11.62   &-2740         &-2450      &1.9:   &$<$1   &1.61      \\     
15080     &44870.5  &0.85   &11.57   &-2730         &-2130      &0.7:   &1.5    &1.28      \\    
          & \nodata &\nodata&\nodata &\nodata       &-2650      &\nodata&\nodata&\nodata   \\
37759     &47867.2  &0.39   &11.36   &-2520         & -2210     & 11.2  & 1.    &1.59      \\       
          &\nodata  &\nodata&\nodata &-2080\tablenotemark{8}     & \nodata   &\nodata&\nodata&\nodata   \\
37768     &47868.2  &0.44   &11.27   &-2550         &-2050      & 10.7  & $<$1  &1.62      \\    
          &\nodata  &\nodata&\nodata &-1890\tablenotemark{8}         & \nodata   &\nodata&\nodata&\nodata   \\
37781     &47870.2  &0.55   &11.27   &-2040         &-l750      & 11.7  & $<$1  &1.86      \\          
37788     &47871.4  &0.61   &11.25   &-2000         &-1950      & 14.3  & 0.2    &1.85      \\       
42446     &48511.5  &0.83   &11.41   &-1900         &-1980      & 12.5  &  $<$1 &1.86      \\        
42470     &48515.6  &0.05   &11.35   &-1700         &-1800      & 16.9  & $<$1  &1.82      \\       
42694     &48541.4  &0.39   &11.55   &-2650         &-2450      & 13.2  &  1.   &1.73      \\     
42702     &48542.4  &0.44   &11.50   &-2480         &-2340      & 11.1  &  2.   &1.86      \\     
42711     &48543.4  &0.49   &11.27   &-2480         &-2210      & 14.5  &  1.4  &1.93      \\   
42721     &48544.4  &0.54   &11.33   &-2450         &-2180\tablenotemark{6}   & 17.8 &  $<$1  &1.95      \\ 
52888                       &49680.5  &0.51   &11.37    &-420         &\nodata    & 7.4   &18.8   &3.32      \\
53036\tablenotemark{10}     &49697.5  &0.39   &\nodata    &-370         &\nodata  &14.0:  &39.1:  &1.96::    \\   
53129\tablenotemark{10}     &49706.5  &0.86   &\nodata    &-470         &\nodata  &6.9:   &34.6:  &1.75::    \\   
53216\tablenotemark{10}     &49716.5  &0.38   &\nodata    &-370         &-1680    &23.9   &57.0   &2.55:      \\  
53226\tablenotemark{10}     &49717.5  &0.43   &11.12\tablenotemark{7}  &-370         &-1730   & 21.9   &58.1   &2.55:     \\ 
54064     &49784.1  &0.89   &\nodata  &-1100          &-2010   & 38.8   &39.3   &2.50      \\  
54490     &49831.0  &0.32   &11.96    &-800\tablenotemark{8}           &\nodata  &43.6   &35.9   &2.65      \\     
          &\nodata  &\nodata&\nodata & -1900        & -1660     &\nodata&\nodata&\nodata   \\
54671     &49851.0  &0.36   &11.98    &-810\tablenotemark{8}           &-950:    &49.0   &35.7   &2.49      \\   
          &\nodata  &\nodata&\nodata & -2000        &-1900      &\nodata&\nodata&\nodata   \\
54727     &49859.9  &0.82   &11.96    &-1100\tablenotemark{8}          & -900    &37.0   &31.6   &2.81      \\       
          &\nodata  &\nodata&\nodata & -2400        &-2300      &\nodata&\nodata&\nodata   \\
55315     &49916.8  &0.78   &11.27    &-1100\tablenotemark{8}          &-800:    &35.9   &20.1   &2.55      \\      
          &\nodata  &\nodata&\nodata & -2530        &-2150      &\nodata&\nodata&\nodata   \\
55380     &49928.8  &0.40   &11.34    &-900\tablenotemark{8}           &-1130    &38.2   &19.2   &2.36      \\     
          &\nodata  &\nodata&\nodata & -2230        & -2400     &\nodata&\nodata&\nodata   \\
55394     &49930.8  &0.507  &11.22    &-900\tablenotemark{8}           &-900:    &38.6   &18.6   &2.38      \\   
          &\nodata  &\nodata&\nodata & -2380        &-2180      &\nodata&\nodata&\nodata   \\
55932     &49975.6  &0.83   &11.22    &-1200\tablenotemark{8}          &-1000    &31.9   &14.7   &2.43      \\
          &\nodata  &\nodata&\nodata & -2620        & -2400:    &\nodata&\nodata&\nodata   \\
55955     &49978.7  &0.99   &11.37    &-1200\tablenotemark{8}         &-880:     &31.9  &14.4    &1.98      \\      
          &\nodata  &\nodata&\nodata & -2530        & -2330     &\nodata&\nodata&\nodata   \\
56017     &49986.6  &0.40   &11.31    &-2670        & -2560      &37.4  &14.4    &2.15      \\     
56188     &50033.5  &0.83   &\nodata  &-1400        &-1850      &29.0   &10.6   &2.29      \\      
56205     &50036.4  &0.99   &\nodata  &-1500        &-1000:     &31.9   & 8.2   &1.76      \\   
          &\nodata  &\nodata&\nodata & \nodata      & -2160     &\nodata&\nodata&\nodata   \\
56223     &50045.2  &0.43   &\nodata  &-2530        & -2500     &30.7   &7.1    &2.15      \\  
  1070    &51304.8  &0.83   &\nodata &-1770:        &-1600     &24.9    &0.8   &2.08      \\              
          &\nodata  &\nodata&\nodata &\nodata        &-2050     &\nodata&\nodata&\nodata   \\
  3070    &51308.9  &0.05   &\nodata  &-1380         &-1570     &25.0   &0.4    &1.68      \\       
  4070    &51310.8  &0.15   &\nodata  &-1350         &-1670\tablenotemark{8} &25.0   &0.6    &2.15      \\      
  5070    &51314.9  &0.36   &\nodata  &-2500         &-1630     &23.6   &0.5    &1.67      \\        
          &\nodata  &\nodata&\nodata &\nodata        &-2130\tablenotemark{9} &\nodata&\nodata&\nodata   \\
  6070    &51315.8  &0.40   &\nodata  &-2140         &-1650     &26.4   &0.2    &1.85      \\            
          &\nodata  &\nodata&\nodata &\nodata        &-2140     &\nodata&\nodata&\nodata   \\
  2070    &51655.1  &0.00   &\nodata  &-1500         &-1830\tablenotemark{8} &17.9   &0.9    &1.68      \\        
  7070    &52386.6  &0.99   & \tablenotemark{11} &-1860        &-2100     &12.2   &\nodata&\nodata   \\ 
  8070    &55083.9  &0.99   &11.6\tablenotemark{12}&-2260      &-2460     & 7.0   &$<$0.2 &1.50      \\       
          &\nodata  &\nodata&\nodata               & -2610     & \nodata  &\nodata&\nodata&\nodata   \\
\enddata
\tablenotetext{1}{Heliocentric Julian Date $-$2400000}
\tablenotetext{2}{Location where P Cyg absorption intersects the continuum level in km/s. 
Corrected for an adopted SMC systemic velocity of $+$150 km/s }
\tablenotetext{3}{Maximum extent of the plateau (``flat" region) of P Cyg absorption in km/s; two values in this
column indicate the presence of a second plateau at a different intensity level. Corrected for an adopted SMC 
systemic velocity of $+$150 km/s}
\tablenotetext{4}{Integrated un-dereddened absolute flux in units of 10$^{-12}$ ergs/cm$^2$-s}
\tablenotetext{5}{Un-dereddened absolute flux in units of 10$^{-12}$ ergs/cm$^2$-s-\AA}
\tablenotetext{6}{The absorption is not flat, but contains considerable structure}
\tablenotetext{7}{Visual magnitude from Koenigsberger et al. 1998b}            
\tablenotetext{8}{Maximum extent of ``plateau"}                                
\tablenotetext{9}{Speed of what may be interpreted as a  short plateau; absorption continues                   
to rise gradually out -3100 km/s }
\tablenotetext{10}{Flux calibration problem}              
\tablenotetext{11}{ Visual magnitude of 11.3 for this same epoch but out of eclipse was provided by      
S. Dufau \& M.T. Ruiz, Private Communication, (2002)}
\tablenotetext{12}{All Sky Automated Survey (ASAS) (Pojmanski, 2002) value obtained $\sim$3 hours earlier}      
\end{deluxetable}

\begin{deluxetable}{lllllllll}
\tablecaption{Low resolution IUE spectra\label{tab_uvspec2} } 
\tablecolumns{7}                  
\tablewidth{0pt}
\tablehead{
\colhead{SWP} & \colhead{HJD$^1$} & \colhead{Phase} & \colhead{FES mag} & \colhead{F(N IV])$^2$} &\colhead{F(NIII)$^2$} &\colhead{F(1850\AA)$^3$}}
\startdata
 1598 &43650.5 & 0.53    &11.72    &   1.8  &  6.52 & 1.61              \\
14112 &44754.4 & 0.82    &11.61    &  1.8   &  2.8  & 1.73                    \\
14135 &44756.2 & 0.92    &11.63    &  5.1   &  0.7  & 1.67                    \\
14166 &44758.8 & 0.05    &11.56    &  4.6   &  1.6  & 1.68                    \\
29633 &46743.7 & 0.07    &11.42    &  5.3   &  0    & 1.69                   \\
29673 &46748.9 & 0.34    &11.59    &  8.9   &  2    & 1.35                    \\
29674 &46748.9 & 0.35    &11.57    &  7.9   &  2    & 1.39                     \\
29681 &46749.7 & 0.39    &11.49    &  6.1   &  2    & 1.52                 \\
29690 &46750.8 & 0.45    &11.44    & 14.7   &  2    & 1.68                 \\
29693 &46750.9 & 0.45    &11.45    &  8.2   &  2    & 1.75                \\
29699 &46751.8 & 0.51    &11.41    &  8.2   &  0    & 1.76                \\
29702 &46751.9 & 0.51    &11.39    &  5.0   &  1    & 1.75                 \\
29705 &46752.7 & 0.54    &11.41    &  7.7   &  2    & 1.77                 \\
29708 &46752.8 & 0.55    &11.39    &  5.3   &  1.5  & 1.76                  \\
29736 &46757.9 & 0.81    &11.37    &  5.9   &  2    & 1.67                 \\
29743 &46758.7 & 0.85    &11.40    &  7.3   &  2    & 1.69                  \\
37760 &47867.5 & 0.39    &11.30    & 11.7   & 2.0   & 1.81               \\ 
37769 &47868.5 & 0.44    &11.25    & 12.8   & 1     & 1.89                \\
37782 &47870.5 & 0.56    &11.22    & 14.0   &  1    & 1.93                \\
37789 &47871.5 & 0.61    &11.20    & 14.3   &  1    & 1.96               \\
37790 &47871.5 & 0.61    &11.37    & 18.6   &  3    & 1.91              \\             
42445 &48511.5 & 0.83    &11.26    & 18.7  &0.9     & 1.99          \\ 
42469 &48515.5 & 0.04    &11.43    & 16.9  &0.5     & 1.82          \\
42693 &48541.4 & 0.38    &12.00    & 12.9  &1       & 1.73          \\
42701 &48542.4 & 0.44    &11.39    & 13.7  &2       & 1.98          \\
42703 &48542.7 & 0.45    &11.49    & 17.2  &0.4     & 2.01          \\
42710 &48543.4 & 0.49    &11.27    & 18.3  &2.8     & 2.06           \\
42720 &48544.4 & 0.54    &11.26    & 14.0  &1.      & 2.04           \\
42722 &48544.7 & 0.55    &11.33    & 16.1  &1.7     & 2.09         \\
52824 &49674.9 & 0.22    &10.11    &  4.0  &26.5    & 3.50        \\
52889 &49680.6 & 0.52    &10.19    &  1.   &27.2    & 3.50        \\
52922 &49684.8 & 0.73    &10.10    &  4    &33.6    & 3.10        \\
52956 &49688.6 & 0.93    &10.22    &   0.8 & 29.3   &  3.05          \\
52967 &49689.7 & 0.99    & \nodata &  10.8 & 37.4   &  3.17             \\
52975 &49690.7 & 0.04    & \nodata &  12.1 & 38.3   &  3.08      \\
52992 &49692.7 & 0.14    &10.54    &  12.0 & 41.8   &  2.74            \\
53035 &49697.2 & 0.38    & \nodata &  23.1 & 42.6   &  2.65         \\
53037 &49697.6 & 0.40    & \nodata &  15.4 & 44.1   &  2.60        \\
53061 &49702.6 & 0.66    & \nodata &  17.1 & 52.3   &  2.65      \\
53128 &49706.3 & 0.85    & \nodata &  18.6 & 45.0   &  2.72       \\
53164 &49709.9 & 0.03    & \nodata &  18.1 & 45.8   &  2.65     \\
53186 &49712.6 & 0.18    & \nodata &  30.7 & 51.5   &  2.80     \\
53187 &49712.7 & 0.18    & \nodata &  31.1 & 53.3   &  2.84      \\
53218 &49716.6 & 0.39    & \nodata &  34.6 & 56.2   &  2.90      \\       
53230 &49717.8 & 0.44    & \nodata &  36.5 & 59.8   &  3.12      \\
54485 &49830.3 & 0.28    & \nodata &  45.8 & 37.1   &  2.89       \\
54486 &49830.5 & 0.29    & \nodata &  40.2 & 37.7   &  2.94     \\
54491 &49831.2 & 0.33    & \nodata &  43.9 & 38.2   &  2.65      \\
54533 &49836.3 & 0.59    & \nodata &  39.8 & 38.3   &  3.24        \\
54534 &49836.3 & 0.59    & \nodata &  43.3 & 42.7   &  3.30      \\
54535 &49836.3 & 0.59    & \nodata &  43.4 &41.9    &  3.23       \\
54664 &49850.3 & 0.32    & \nodata &  41.9 &38.6    &  2.62       \\
54665 &49850.4 & 0.33    & \nodata &  42.1 &38.1    &  2.68      \\
54666 &49850.4 & 0.33    & \nodata &  46.4 &36.1    &  2.70       \\
54670 &49850.8 & 0.35    & \nodata &  47.4 &39.5    &  2.61     \\
54708 &49857.4 & 0.69    & \nodata &  44.0 &37.8    &  2.85                      \\
54709 &49857.4 & 0.69    & \nodata &  40.4 &37.2    &  2.93                      \\
54728 &49860.1 & 0.83    & \nodata &  45.5 &30.1    &  2.93                     \\
54758 &49864.1 & 0.04    & \nodata &  37.7 &34.0    &  2.53                      \\
54759 &49864.2 & 0.04    & \nodata &  35.1 &34.4    &  2.37                      \\
54760 &49864.2 & 0.04    & \nodata &  32.6 &37.8    &  2.26                      \\
55111 &49894.1 & 0.59    & 10.94   &  41.6 &27.0    &  2.60                      \\
55112 &49894.1 & 0.60    & 10.21   &  45.3 &29.5    &  2.65                      \\
55314 &49916.7 & 0.77    & 11.27   &  39.3 &20.4    &  2.69                     \\
55340 &49921.0 & 0.99    & \nodata &  41.2 &23.8    &  2.33                      \\
55381 &49928.9 & 0.40    &  11.30  &  42.0 &20.9    &  2.38                      \\
55382 &49928.9 & 0.51    & \nodata &  40.2 &21.6    &  2.48                      \\
55382 &49928.9 & 0.51    & \nodata &  40.2 &21.6    &  2.48                      \\
55395 &49930.9 & 0.51    & 11.19   &  42.5 &24.1    &  2.60                      \\
55396 &49930.9 & 0.51    & 11.25   &  40.3 &22.1    &  2.55                      \\
55461 &49939.9 & 0.97    & 11.33   &  43.7 &24.2    &  2.17                      \\
55462 &49939.9 & 0.98    & \nodata &  39.0 &18.3    &  2.13                      \\
55931 &49975.5 & 0.82    & \nodata &  45.5 &13.0    &  2.48                      \\
55933 &49975.8 & 0.84    & \nodata &  40.4 &15.3    &  2.40                      \\
55954 &49978.8 & 0.99    & \nodata &  41.9 &19.0    &  1.96                      \\
55956 &49978.8 & 0.99    & \nodata &  36.8 &14.9    &  2.14                      \\
55957 &49978.8 & 0.00    & \nodata &  37.1 &14.6    &  2.07                     \\       
55958 &49978.9 & 0.00    & \nodata &  35.5 &18.6    &  2.08                      \\
55976 &49981.8 & 0.15    & \nodata &  33.0 &15.8    &  2.40                      \\
55977 &49981.9 & 0.15    & \nodata &  39.2 &18.1    &  2.48                      \\
56005 &49985.0 & 0.32    & \nodata &  42.7 &17.3    &  2.03                      \\
56006 &49985.1 & 0.33    & \nodata &  40.0 &13.9    &  2.20                      \\
56013 &49985.9 & 0.36    & \nodata &  37.2 &16.3    &  2.17                      \\
56014 &49985.9 & 0.36    & \nodata &  36.7 &13.0    &  2.08                      \\
56015 &49985.9 & 0.36    & \nodata &  32.0 &13.2    &  2.13                      \\
56018 &49986.7 & 0.40    & \nodata &  42.2 &15.3    &  2.34                      \\
56033 &49990.8 & 0.62    & \nodata &  38.7 &15.3    &  2.48                      \\
56034 &49990.9 & 0.62    & \nodata &  31.7 &14.3    &  2.47                      \\
\enddata
\tablenotetext{1}{Heliocentric Julian Date $-$2400000}
\tablenotetext{2}{Integrated un-dereddened absolute flux in units of 10$^{-12}$ ergs/cm$^2$-s}
\tablenotetext{3}{Un-dereddened absolute flux in units of 10$^{-12}$ ergs/cm$^2$-s-\AA}
\tablenotetext{d}{Visual magnitude from Koenigsberger et al. 1998b}
\end{deluxetable}

\begin{deluxetable}{lrcccl}
\tablecaption{Estimated wind speeds\label{tab_star_velocities} }
\tablecolumns{6}
\tablewidth{0pt}
\tablehead{
\colhead{Year} & \multicolumn{3}{c}{}&\colhead{ }  & \colhead{$V_{edge}$ (HeII)}  \\
\colhead{    } & \colhead{{\it Star A}}&\colhead{{\it Star B}}&\colhead{{\it Star C}}&\colhead{ }&\colhead{{\it Star A}}
}
\startdata
1979       &     $-$2670\tablenotemark{1}         &  \nodata       & $-$1760  &     &$-$2560    \\
1991       &     $-$2000\tablenotemark{1}         &$-$2320         & $-$1740  &     &$-$1870   \\
1993.7     &     $-$1610\tablenotemark{2}         &  \nodata       &  \nodata &     &\nodata \\
1994.90    &     $-$690\tablenotemark{3}          &  \nodata       &  \nodata &     &\nodata \\
1994.94    &     $-$810\tablenotemark{3}          &  \nodata       &  \nodata &     &\nodata \\
1994.97    &     $-$860\tablenotemark{3}          &  \nodata       &  \nodata &     &\nodata \\
1994.98    &     $-$1100\tablenotemark{3}         &  \nodata       & \nodata  &     &\nodata \\
1995.00    &     $-$1300\tablenotemark{3}         &  \nodata       &  \nodata &     &\nodata \\
1999       &     $-$1720\tablenotemark{1}         &$-$2300         & $-$1740  &     &$-$1500   \\
2000       &     $-$2000\tablenotemark{1}         &  \nodata       & $-$1780  &     &$-$1680   \\
2002       &     $-$2210\tablenotemark{1}         &  \nodata       & $-$1780  &     &$-$2100   \\
2009       &     $-$2440\tablenotemark{1}         &  \nodata       & $-$1760  &     &$-$2260   \\
\enddata
\tablenotetext{1}{From $V_{black}$ C IV}
\tablenotetext{2}{From P V, {\it FUSE}, Koenigsberger et al. 2006}   
\tablenotetext{3}{From Al III, Koenigsberger et al. 1998a}
\end{deluxetable}

\begin{deluxetable}{lccccccccc}
\rotate
\tablecaption{Ions and number of levels treated in the model \label{table_atoms}}
\tablecolumns{8}
\tablehead{ \colhead{Elements}    & \colhead{I}  &\colhead{II}     &\colhead{III}  &\colhead{IV} &\colhead{V} &\colhead{VI} &\colhead{VII} &\colhead{VIII} &\colhead{IX} }
\startdata
H      &  20/30      &      &       &        &       &     &      &      &       \\
He    &  27/39     &  13/30    &       &        &       &     &      &      &       \\
C      &        &   40/92   &  51/84     &   59/64     &       &     &      &      &       \\
N      &       &   45/85   &  41/82     &   44/76     &   41/49    &     &      &      &       \\
O      &       &   54/123   &   88/170    &   38/78     &   32/56    &  25/31   &      &      &       \\
Ne      &       &   42/242   &   40/182    &   53/355     &  37/166     &  36/202   &      &      &       \\
Al      &       &      &   21/65    &        &       &     &      &      &       \\
Si      &       &      &   20/34    &   22/33     &       &     &      &      &       \\
P       &       &      &       &   36/178     &   12/62    &     &      &      &       \\
S       &       &      &   13/28    &   51/142     &   31/98    &  28/58   &      &      &       \\
Cl      &       &      &       &    40/129    &   26/80    &   18/44  &   17/28   &      &       \\
Ar      &       &      &  10/36     &   31/105     &   38/99    &     &      &      &       \\
Ca     &       &      &       &    43/378    &   73/613    &  47/108   &   48/288   &  45/296    &   39/162    \\
Cr      &       &      &    30/145   &   29/234     &   30/223    &  30/215   &      &      &       \\
Mn      &       &      &       &   39/464     &   16/80    &   23/181  &   20/203   &      &       \\
Fe     &       &      &   104/1433    &   100/1000     &   139/1000    &  44/433   &  29/153    &      &       \\
Ni      &       &      &       &    115/1000    &   152/1000    &  62/1000   &   37/308   &   34/325   &   34/363    \\
\enddata
\end{deluxetable}

\begin{deluxetable}{lcccccc}

\tablewidth{0pt}
\tablecaption{Chemical composition of {\it Star A} compared to other objects\label{table_smc_comp}}
\tablecolumns{7}
\tablehead{\colhead{Element} &\colhead{{\it Star A}\tablenotemark{a}} &\colhead{{\it Star C} \tablenotemark{a}} 
&\colhead{SMC\tablenotemark{b}}&\colhead{NGC346\tablenotemark{c}}& \colhead{AG Car \tablenotemark{e}} &   
\colhead{Sun \tablenotemark{d}} }
\startdata
H     &     11.304 &    11.853 &    \nodata &    \nodata &     11.868 &    11.868 \\
He    &     11.902 &    11.455 &    \nodata &     11.381 &     12.097 &    11.397 \\
C     &      7.484 &     8.233 &    \nodata &    \nodata &      8.694 &     9.334 \\
N     &      9.000 &     7.999 &      7.526 &      7.521 &     10.161 &     8.791 \\
O     &      6.176 &     9.057 &      9.224 &      9.219 &      8.679 &     9.729 \\
Ne    &      8.241 &     8.542 &      8.474 &      8.470 &    \nodata &     9.010 \\
Na    &      6.538 &     6.839 &    \nodata &    \nodata &    \nodata &     7.396 \\
Mg    &      7.811 &     8.111 &    \nodata &    \nodata &    \nodata &     8.780 \\
Al    &      6.747 &     7.049 &    \nodata &    \nodata &      8.056 &     7.666 \\
Si    &      7.844 &     8.146 &    \nodata &    \nodata &      9.123 &     8.823 \\
P     &      5.787 &     6.086 &    \nodata &    \nodata &    \nodata &     6.716 \\
S     &      7.562 &     7.863 &      7.966 &      7.961 &    \nodata &     8.511 \\
Cl    &      5.896 &     6.196 &    \nodata &    \nodata &    \nodata &     6.914 \\
Ar    &      6.009 &     7.310 &      7.291 &      7.286 &    \nodata &     7.646 \\
Ca    &      6.809 &     5.865 &    \nodata &    \nodata &    \nodata &     7.778 \\
Ti    &      5.516 &     7.090 &    \nodata &    \nodata &    \nodata &     6.445 \\
Cr    &      6.230 &     6.531 &    \nodata &    \nodata &    \nodata &     7.221 \\
Mn    &      5.975 &     6.274 &    \nodata &    \nodata &    \nodata &     6.995 \\
Fe    &      8.134 &     8.436 &    \nodata &    \nodata &      9.442 &     9.062 \\
Ni    &      6.865 &     7.164 &    \nodata &    \nodata &    \nodata &     7.863 \\

\enddata   

\tablenotetext{a}{This paper;}
\tablenotetext{b}{Hunter et al. 2007;}   
\tablenotetext{c}{Peimbert et al. 2000;}  
\tablenotetext{d}{Grevesse et al.2007;}   
\tablenotetext{e}{Groh et al., 2009;}   
\end{deluxetable}

\begin{deluxetable}{lccccccc}

\tablewidth{0pt}
\tablecaption{Model Fit Results\label{tab_params}}
\tablecolumns{8}
\tablehead{\colhead{Parameter} &\colhead{Sept. 1994\tablenotemark{a}} &\colhead{Dec. 1994\tablenotemark{b}} &\colhead{Dec. 1994\tablenotemark{c} }     &\colhead{2000\tablenotemark{c}} & \colhead{2002\tablenotemark{c}} & \colhead{2009\tablenotemark{c}} & \colhead{starC\tablenotemark{c}}}
\startdata
JD - 240000.0 &  \nodata    &    49716.6   &   49716.6  & 51655.1 & 52386.6 & 55083.9 &  \nodata \\
V [mag] (system)  &  \nodata    &    11.12     &   11.12 \tablenotemark{b}   & 11.3  &  11.6 &  \nodata & \nodata      \\
R$_{10}$  [$R_\odot$]                                           &   \nodata   &  48 \tablenotemark{d}     &   28        &  20.4        &  21       &   19.3     &   23.5 \\
R$_s$    [$R_\odot$]                                         &  \nodata    & \nodata                                &   21.5     &   24.3   &    21.2   &   19.6    &   25.0 \\
R$_{2/3}$  [$R_\odot$]                                     &   280          &\nodata                                 &   124      &  34        &  32      &   28     &   24.2 \\
$\tau_s$                                                               & \nodata     & \nodata                                &    25.8    &  4.7      &  3.7       & 1.8   & 0.005 \\
T$_{eff}$  [kK]                                                      &   \nodata   &\nodata                                &   23         &  37.3     &  40       &   43     &   32 \\
T$_*$  [kK]                                                            &  23             &   35.5 \tablenotemark{d}  &   47        &  48        &  50       &   47      &   33 \\
T$_s$ [kK]                                                             &  \nodata    & \nodata                               & 100       &  57       &   58        &   60     &    27 \\
$\dot{M}/\sqrt{f}$ [10$^{-5}$M$_\odot/yr$]      &  80             & 100                                      &   111     &  35        &  25     &    23   &   0.06  \\
Log(L/L$_\odot$)                                                &  7.05          & 6.48                                     &   6.57    &  6.30     &  6.39    &   6.39     &   5.77 \\
V$_\infty$  [km/s]                                                 &   500          & 600                                      &   750     &  2000    &  2200   &   2440   &  1800 \\
V$_{esc}$ [km/s]                                                  &     464       &            \nodata                     &   642      &    832      &  895      &     932      &    \nodata   \\
$\Gamma$                                                            &      0.75\tablenotemark{e}   &  \nodata  &   0.75   &     0.53    &   0.53      &    0.53     &  \nodata   \\
\enddata

\tablenotetext{a}{Drissen et al. 2001;} 
\tablenotetext{b}{Koenigsberger et al. 1998b;}
\tablenotetext{c}{This paper;}
\tablenotetext{d}{``Core" radius, R$_*$, and corresponding temperature; in Koenigsberger et al. (1998b), R$_*$ is  
the inner boundary of the model atmosphere, where the expansion velocity is negligible}
\tablenotetext{e}{Adopted value same as in 1994 model.}
\end{deluxetable}

\newpage
\clearpage\eject

\clearpage

\end{document}